\documentclass[journal,onecolumn,11pt]{IEEEtran}
\usepackage{cite,graphicx,amsmath,amssymb,eufrak,mathrsfs,epsf,epsfig}
\usepackage[usenames]{color}
\usepackage{psfrag}

\usepackage{multirow,subcaption}

\usepackage[active]{srcltx}

\usepackage{algorithm}
\usepackage{algpseudocode}

\ifCLASSINFOpdf

\else
\fi

\begin{document}
\newtheorem{ach}{Achievability}
\newtheorem{con}{Converse}
\newtheorem{definition}{Definition}
\newtheorem{theorem}{Theorem}
\newtheorem{lemma}{Lemma}
\newtheorem{example}{Example}
\newtheorem{cor}{Corollary}
\newtheorem{prop}{Proposition}
\newtheorem{conjecture}{Conjecture}
\newtheorem{remark}{Remark}
\title{Dynamic Multichannel Access via Multi-agent Reinforcement Learning: Throughput and Fairness Guarantees}
\author{Muhammad Sohaib, Jongjin Jeong, and Sang-Woon Jeon
\thanks{M. Sohaib is with the Department of Electronics Engineering, Hanyang University, Ansan 15588, South Korea (e-mail: msohaib828@hanyang.ac.kr).}
\thanks{J. Jeong is with the Department of Electronics Engineering, Hanyang University, Ansan 15588, South Korea (e-mail: jeongjj@hanyang.ac.kr).}
\thanks{S.-W. Jeon, corresponding author, is with the Department of Military Information Engineering, Hanyang University, Ansan 15588, South Korea (e-mail: sangwoonjeon@hanyang.ac.kr).}%
}%
 \maketitle

\begin{abstract}
We consider a multichannel random access system in which each user accesses a single channel at each time slot to communicate with an access point (AP). Users arrive to the system at random and be activated for a certain period of time slots and then disappear from the system. Under such dynamic network environment, we propose a distributed multichannel access protocol based on multi-agent reinforcement learning (RL) to improve both throughput and fairness between active users. 
Unlike the previous approaches adjusting channel access probabilities at each time slot, the proposed RL algorithm deterministically selects a set of channel access policies for several consecutive time slots.
To effectively reduce the complexity of the proposed RL algorithm, we adopt a branching dueling Q-network architecture and propose an efficient training methodology for producing proper Q-values over time-varying user sets. We perform extensive simulations on realistic traffic environments and demonstrate that the proposed online learning improves both throughput and fairness compared to the conventional RL approaches and centralized scheduling policies.
\end{abstract}
 \IEEEpeerreviewmaketitle



\section{Introduction}  \label{intro}
In order to satisfy soaring data demands with scarce and costly wireless communication resources, real-time resource allocation and management techniques adjusting dynamic network environment have been actively studied for fifth-generation (5G) and beyond 5G cellular systems~\cite{12,22}.
The core of such techniques is to provide flexible and adaptive resource access policies for end terminals in a distributed and decentralized manner, which is essentially required for internet-of-things (IoT) or machine-type applications~\cite{41,42,9260174}. 
Distributed resource or channel access is also crucially important to provide massive connectivity for such applications.

Traditionally, random access has been considered as a promising solution for efficiently sharing wireless resource between multiple terminals without coordination~\cite{43,45}. It has been known that the tree (or splitting) algorithm \cite{43} and its variants \cite{45, 46} provide improved throughputs of ALOHA-based random access, which yields the maximum throughput of  $0.487$ (packets/slot).
Later, successive interference cancellation (SIC) techniques have been incorporated in ALOHA-based random access, achieving the maximum throughput of $0.693$ for single packet transmission \cite{47, jeon} and the maximum throughput close to one for multiple packet transmission \cite{48}.

Even through introducing SIC receivers can drastically improve the sum throughput of ALOHA random access systems \cite{47,jeon, 48}, implementing such sophisticated physical layer technique is quite challenging in practice due to signaling overhead and channel estimation error. To overcome this limitation, reinforcement learning (RL) has been recently adopted in the literature to provide an enhanced throughput by avoiding collision between terminals. 
In \cite{CHU201523}, RL has been applied to ALOHA systems in order to improve throughput and energy efficiency of single-hop wireless sensor networks.
RL algorithms have been applied for multichannel access problems to identify correlated channel patterns, which were modeled by Markov processes, and also avoid collisions from other transmitted terminals in \cite{17,28}.  
A more general channel model yielding different expected rates between terminals has been considered in \cite{3}.
Various RL algorithms have been further proposed for adjusting dynamic environment for vehicular networks \cite{26,27} and for jointly optimizing user association and resource allocation for heterogeneous multi-cell networks \cite{6,38}.
In \cite{21}, a two-stage RL approach for a single-user scenario has been proposed for channel sensing and selection in cognitive radio networks, which leads to higher throughput with lower energy consumption.
Similarly in \cite{20,34}, spectrum sharing strategies have been developed for cognitive radios, where primary and secondary users operate in a non-cooperative manner.
Also, with non-orthogonal multiple access (NOMA) being considered as an important technique for improving throughput and supporting massive connectivity in next-generation wireless networks, various RL-based channel assignment techniques have been developed to improve the quality of service (QoS) of NOMA-based systems \cite{9352956,9102308,8952876}.

Note that most existing works for multichannel access via multi-agent RL focused on throughput or QoS maximization under a certain set of constraints \cite{17,28,3,26,27,6,38,23}. 
On the other hand, guaranteeing fairness between accessed terminals is required for various random access systems \cite{8,9}. 
To meet such requirement, proportional fairness (PF) has been recently considered to provide fair resource assignment for multi-agent RL frameworks \cite{40, 11, 25}.  

In this paper, we propose a multi-agent distributed multichannel access policy guaranteeing fair multichannel access between accessed terminals.
Unlike the previous multichannel access approaches \cite{43,17,28,11}, in which each RL agent controls multichannel access probabilities at each time slot for minimizing collision  \cite{43,17,28} or providing PF between terminals \cite{11}, we adopt a deterministic multichannel access policy in which each RL agent determines a set of multichannel access policies for several consecutive time slots to enhance both throughput and fairness. More importantly, we focus on dynamic network environments in which each terminal arrives at random and be only activated for a finite number of time slots and, as a consequence, the number of active terminals in the system varies over time.
Therefore, guaranteeing real-time fairness while minimizing collisions between active terminals becomes quite challenging.
We highlight our contributions in this work as follows:
\begin{itemize}
\item We propose an online learning framework and related learning methodologies applicable for dynamic network environments that are able to provide fair multichannel access between active terminals in real-time by adjusting their transmission policies depending on the number of active users in a fully distributed and decentralized manner. 
\item Because each active terminal is required to jointly establish its multichannel access policy for multiple time slots, the system complexity increases linearly as the average number of active terminals increases. To provide a scalable multichannel access solution for massive terminal systems, we propose a vectorized Q-value evaluation and the related action construction methods in which their computational complexity increase marginally for an increasing number of active terminals.
\item We further generalize the proposed online learning framework applicable for general time-varying wireless channels and rate-based throughput metrics. The performance evaluations are carried out with the ‘Simulation of Urban MObility (SUMO) simulator’ package by generating realistic vehicular channel environments. 
\end{itemize}

\subsection{Notation and Paper Organization}

Throughout the paper, $[1:n]$ denotes $\{1,2\cdots, n\}$.
For a vector $\mathbf{a}$, denote the $i$th element of $\mathbf{a}$ by $[\mathbf{a}]_i$.
For a matrix $\mathbf{A}$, denote the $(i,j)$th element of $\mathbf{A}$ by $[\mathbf{A}]_{(i,j)}$.
For a set $\mathcal{A}$, denote the cardinality of $\mathcal{A}$ by $|\mathcal{A}|$.
The indicator function is denoted by $\mathbf{1}(\cdot)$.

The rest of the paper is organized as follows. Section \ref{sec:PF} describes the system model and the performance metrics used throughout the paper. Section \ref{sec:rl} explains the proposed multi-agent reinforcement learning based multichannel access algorithm. In Section \ref{sec:DQN}, the related training methodology for deep Q-networks is provided. Section \ref{sec:result} demonstrates the performance of the proposed scheme by numerical evaluation. Discussion for extending the proposed framework to general wireless environment is given in Section \ref{sec:sumo} and concluding remarks are given in Section \ref{sec:conclusion}.

\section{Problem Formulation} \label{sec:PF}
In this section, we introduce the system model and the performance metrics used throughout the paper.

\subsection{Distributed Multichannel Access}
\begin{figure}[t] \centering
\includegraphics[scale=0.26]{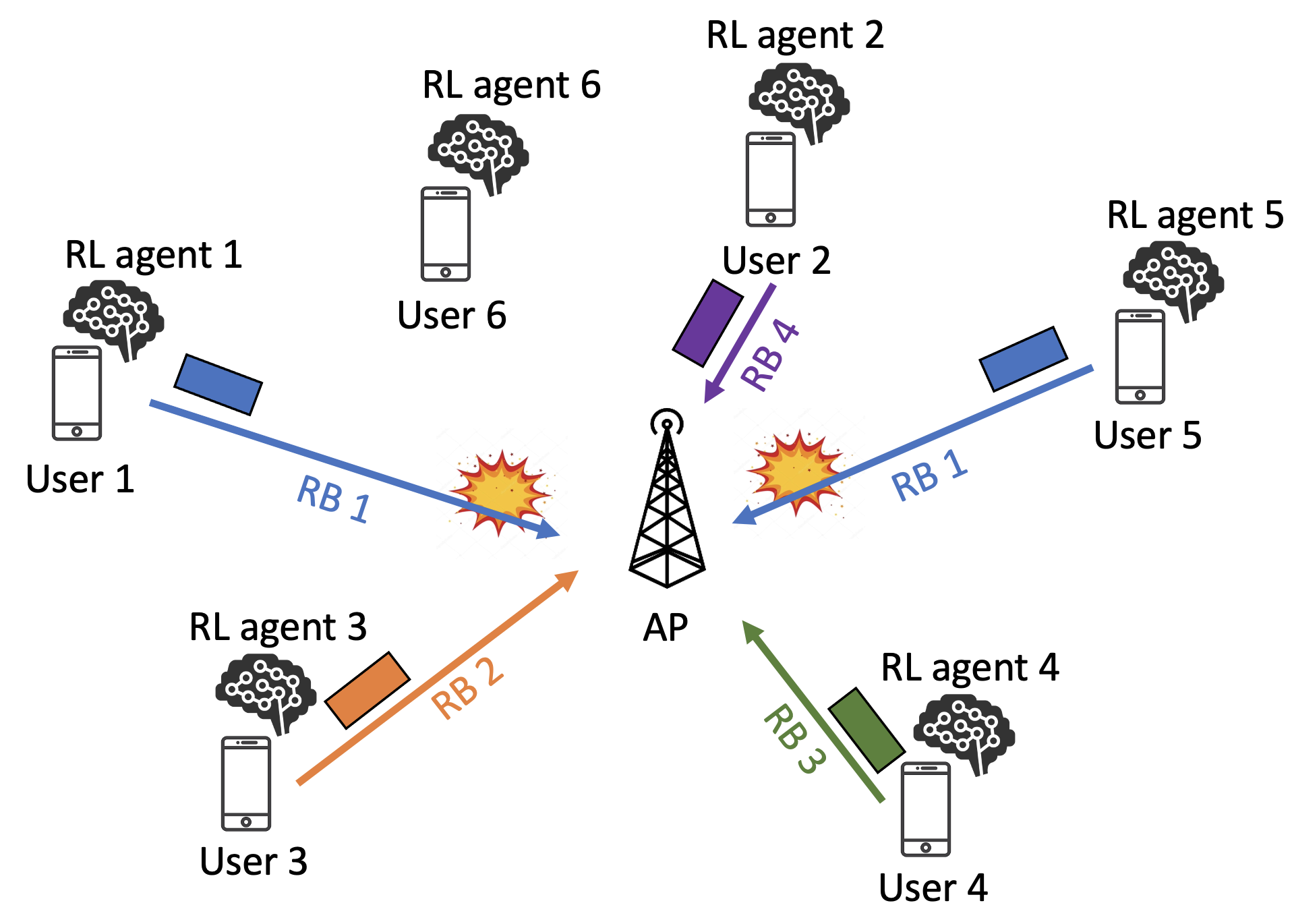}
\caption{Multichannel random access via multi-agent reinforcement learning.}\label{fig:system_model}
\end{figure}

We consider a wireless communication system in which an access point (AP) is located at the center of  a cell and serves users in the cell using $N$ orthogonal resource blocks (RBs) in Fig. \ref{fig:system_model}.
We assume that transmission time is slotted of equal length for one packet transmission. 
Users arrive to the system at random and be activated for a certain period of time slots and then disappear from the system.\footnote{The activation time duration might be different for each user.} 
Due to the random arrival of users, the number of active users in the system varies over time. Such a time-varying active user set is the main difficulty of designing efficient protocols for distributed resource management. 
Let $\mathcal{K}(t)$ be the set of active users at time slot $t$.
At each time slot $t$, a subset of users in $\mathcal{K}(t)$ transmit their packets based on a predetermined distributed protocol, which will be specified later, and consequently the number of packets arrived at the AP will vary over time. For ease of explanation, we remove the corresponding time slots when there is no active user and therefore assume $\mathcal{K}(t)\neq \emptyset$ for all $t$ from now on.

For $k\in\mathcal{K}(t)$, let
\begin{align} \label{eq:eta_k}
\eta_k(t)\in[0:N]
\end{align} 
be the multichannel access indicator of user $k$ at time slot $t$.
That is, user $k$ sends a packet at time slot $t$ through RB $\eta_k(t)$ if $\eta_k(t)\in[1:N]$ and does not send any packet if  $\eta_k(t)=0$, i.e., no transmission. 
When a single packet is received via RB $n$ at the AP, it can decode the corresponding packet successfully. On the other hand, if the AP receives multiple packets simultaneously via RB $n$, it cannot decode the packets, i.e., \emph{collision} occurs. 

At the end of each time slot $t$, the AP broadcasts `ACK' or `NAK' information of all RBs to the users in $\mathcal{K}(t)$. That is, $N$ bits of feedback information is available at each user.
For $n\in[1:N]$, denote the feedback message for RB $n$ at time slot $t$ by $b^{[n]}(t)$, which is set as
\begin{align} \label{eq:ack_nak}
b^{[n]}(t)=\begin{cases}
\operatorname{ACK} &\mbox{ if }\sum_{k\in\mathcal{K}(t)}\mathbf{1}(\eta_k(t)=n)=1, \\
\operatorname{NAK} &\mbox{ otherwise.}
\end{cases}
\end{align}
For notational convenience, denote the set of $N$ bits of feedback messages at time slot $t$ by $\mathbf{b}(t)=\big[b^{[1]}(t),\cdots, b^{[N]}(t)\big]$.

The instantaneous normalized achievable throughput of user $k$ at time slot $t$ is then represented by 
\begin{align} \label{eq:ins_th}
\gamma_k(t)=\begin{cases}\mathbf{1}(b^{[\eta_k(t)]}(t)=\operatorname{ACK}) &\mbox{ if }\eta_k(t)\in[1:N],\\
0 &\mbox{ otherwise.}
\end{cases}
\end{align}
That is, $\gamma_k(t)=1$ if a single packet was successfully transmitted at time slot $t$ and $\gamma_k(t)=0$ for the case of collision or no transmission at time slot $t$.
In Section \ref{sec:sumo}, we will present how to extend our proposed framework for actual rate-based throughput metrics reflecting more realistic time-varying channel environment.

This paper mainly focuses on the regime where $\mathcal{K}(t)\geq N$ and the primary aim is to establish an efficient multichannel access protocol that is able to equally share RBs between active users in $\mathcal{K}(t)$ in a distributed and decentralized manner. For such purpose, we formally define a short-term average target per-user throughput and short-term average achievable per-user throughput in the next subsection.

\subsection{Short-Term Fairness} \label{subset:st_fair}
To measure fairness between the active users in $\mathcal{K}(t)$ in real-time, we first define an instantaneous target per-user throughput at time slot $t$ as\footnote{Since each user can access a single channel at each time  slot, the maximum per-user throughput is limited by one.}
\begin{align} \label{eq:th_target}
\gamma_{\operatorname{target}}(t)=\min\left(1,\frac{N}{|\mathcal{K}(t)|}\right). 
\end{align} 
Denote the arrival and departure time of user $k$ by $t_{\operatorname{arr}, k}$ and $t_{\operatorname{dep},k}$ respectively.
Then a short-term average target per-user throughput of user $k$ at time slot $t$ is defined as
\begin{align} \label{eq:th_target_ave}
\Gamma_{\operatorname{target},k}(t):=\frac{\sum_{i=\max(t_{\operatorname{arr},k},t-T_w)}^{t} \gamma_{\operatorname{target}}(i)}{t-\max(t_{\operatorname{arr},k},t-T_w)+1}
\end{align}
for $t\in[t_{\operatorname{arr}, k}:t_{\operatorname{dep},k}]$, where $T_w>0$ is the average window size, which is the average target per-user throughput (averaged over time slots between $\max(t_{\operatorname{arr},k},t-T_w)$ and $t$).
Similarly,  a short-term achievable throughput of user $k$ at time slot $t$ is defined as
\begin{align}  \label{eq:th_ave}
\Gamma_k(t):=\frac{\sum_{i=\max(t_{\operatorname{arr},k},t-T_w)}^{t}\gamma_k(i)}{t-\max(t_{\operatorname{arr},k},t-T_w)+1}
\end{align}
for $t\in[t_{\operatorname{arr}, k}:t_{\operatorname{dep},k}]$, where $\gamma_k(t)$ is given in \eqref{eq:ins_th}.

Let $T_k=t_{\operatorname{dep},k}-t_{\operatorname{arr}, k}+1$, which is the active time duration of user $k$.
Now, define  a short-term average throughput loss between $\Gamma_{\operatorname{target},k}(t)$ and $\Gamma_k(t)$ during the time slots in $[t_{\operatorname{arr}, k}:t_{\operatorname{dep},k}]$ by
\begin{align} \label{eq:Delta_k}
\Delta_k=\frac{1}{T_k}\sum_{t=t_{\operatorname{arr}, k}}^{t_{\operatorname{dep},k}}\max(\Gamma_{\operatorname{target},k}(t)-\Gamma_k(t),0).
\end{align}
As seen in the definition of $\Delta_k$, there is no advantage of achieving a larger throughput than the target throughput because it may prevent other users from achieving their throughputs close to their target throughputs.

Denote the total number of served users completed up to time slot $t$ by $K_{\operatorname{total}}(t)=\{k| t_{\operatorname{dep},k}\leq t \mbox{ for all }k \in\bigcup_{t'\in[1:t]} \mathcal{K}(t')\}$. 
Also, let $\pi_k$ be a multichannel access policy of user $k$. The aim of this paper is to establish a set of policies $\{\pi_1,\pi_2, \cdots\pi_{K_{\operatorname{total}(t)}}\}$ such that
\begin{align} \label{eq:min_optimization}
\limsup_{t\to\infty}{\arg\min}_{\{\pi_1,\pi_2, \cdots\pi_{K_{\operatorname{total}}(t)}\}}\sum_{k=1}^{K_{\operatorname{total}}(t)} \frac{T_k}{\sum_{l=1}^{K_{\operatorname{total}}(t)} T_l}\Delta_k.
\end{align}
Here, $ \frac{T_k}{\sum_{l=1}^{K_{\operatorname{total}}(t)}T_l}$ is the weight for user $k$, which is proportional to its active time duration $T_k$. 

\section{Multi-Agent Deep Reinforcement Learning} \label{sec:rl}
In this subsection, we propose a multi-agent distributed multichannel access policy guaranteeing fair multichannel access between active users.
For such purpose, we assume that a separate RL agent is implemented at each user. 
For convenience, we simply denote the `agent of user $k$' by `user $k$' from now on.

\subsection{Main Idea and Technical Challenges}
For better understanding, we briefly provide a high level description of the proposed approach and  technical challenges in this subsection.
Traditionally, distributed random access policies have controlled the transmission probability of each user to minimize collision between the transmitted users~\cite{43,45}. However, the throughput of such policies is limited by $0.487N$ achievable by tree-based collision resolve algorithms~\cite{43,45}.
Recently developed RL based channel access and transmission probability adaptation schemes allow much higher throughput by learning historical access information at each user and adjusts its current access policy~\cite{43,44,45,46,17,28,11}. 
To guarantee fairness between users, the PF criteria has been adopted in \cite{11} for adjusting the transmission probability of each user, which provides an improved fairness between users while achieving the sum throughput of approximately $0.8N$.
  
In this paper, we further improve both throughput and fairness, targeting to achieve the sum throughput of $N$ and per-user throughput of $N/K$ when the number of users is given and fixed as $K$ and $K\geq N$. 
More importantly, unlike the previous works \cite{43,44,45,46,17,28,11}, we focus on the dynamic network environment in which each user arrives at random and sends a finite number of packets and, as a consequence, the number of active user in the system varies over time.
Therefore, guaranteeing the real-time fairness as defined in \eqref{eq:Delta_k} and \eqref{eq:min_optimization} without collision becomes quite challenge for such dynamic environment.

To resolve the limitation of the previous approaches \cite{43,44,45,46,17,28,11} which basically adjust transmission probabilities of users, we introduce the deterministic transmission policy over consecutive time slots. 
As a simple example, consider the case where $N=1$ and the number of users is given and fixed as $K$.
If each user can access the channel once at every $K$ time slots without collision, then $1/K$ throughput is achievable at each user.
We notice that such a deterministic random access strategy, i.e, each user is required to deterministically choose its transmission time among $K$ time slots, might helpful for improving both throughput and fairness, because uncertainty by introducing transmission probabilities at each time slot can be removed. 
However, there are two main technical issues to be appropriately  addressed: 
\begin{itemize}
\item Because each active user is required to jointly choose its multichannel access policy for multiple time slots, the system complexity increases exponentially as the average number of active users in each time slot increases. Therefore, to provide a scalable multichannel access solution for massive user systems, algorithm complexity should increase marginally for an increasing number of active users.
\item For the considered dynamic network environment, the number of active users varies over time slots which makes hard to design a universally efficient multichannel access policy. In particular, a systematic RL framework and related learning methodologies that are able to deal with such dynamic environment should be developed.
\end{itemize}

In the following, we describe our proposed multichannel access policy addressing the above technical issues.

\subsection{Multichannel Access Policy} \label{subsec:policy}
\begin{figure}[t] \centering
\includegraphics[scale=0.35]{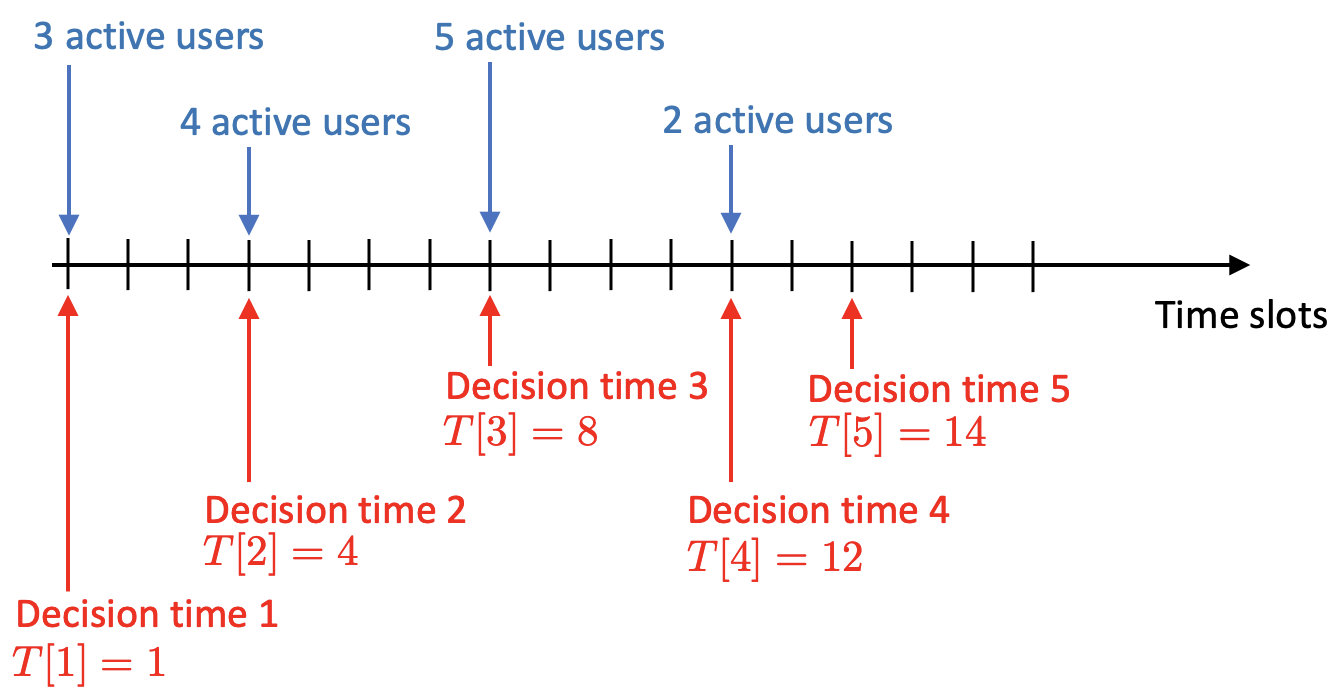}
\caption{Example for decision times when $K_{\max}=4$.}\label{fig:decision_time}
\end{figure}

As mentioned previously, each active user will determine its transmission policy for multiple consecutive time slots.
For convenience, we introduce `decision times' to represent certain time slots for such decision.
To formally introduce decision times, we firstly define a predetermined positive integer value $K_{\max}$. The operational role of $K_{\max}$ will be explained later.
Then define $T[1]=1$ and $T[i]=T[i-1]+\min(|\mathcal{K}(T[i-1])|,K_{\max})$ for $i\geq 2$.
For notational simplicity, denote $K[i]=\min(|\mathcal{K}(T[i])|,K_{\max})$.
At time slot $t=T[i]$, each active user in $\mathcal{K}(T[i])$ determines its transmission policy for the $K[i]$ consecutive time slots, i.e., $t\in[T[i]:T[i]+K[i]-1]$. 
We will use `decision time $i$', which corresponds to time slot $t=T[i]$.
Fig. \ref{fig:decision_time} illustrates an example for decision times when $K_{\max}=4$.
Note that each decision time is determined based on $\{T[i]\}_i$, which are independent of the number of resource blocks $N$.

Recall that user $k$ arrives at time slot $t_{\operatorname{arr}, k}$ and departures at time slot $t_{\operatorname{dep},k}$. Hence, user $k$ can participate in multichannel access during the decision times $[i_{\operatorname{arr}, k}:i_{\operatorname{dep},k}]$, where $i_{\operatorname{arr}, k}$ is the smallest integer $j$ satisfying that $T[j]\geq t_{\operatorname{arr}, k}$ and  $i_{\operatorname{dep}, k}$ is the largest integer $j$ satisfying that $T[j]\leq t_{\operatorname{dep}, k}$.
For each decision time, an RL agent of user $k$ takes its action from the current state information and the corresponding reward will be given.
To formally state the above procedure, let $\mathbf{a}_{k}[i]$, $r_k[i]$, and $\mathbf{s}_k[i]$ be the action, reward, and state of user $k$ at decision time $i$, where $i\in[i_{\operatorname{arr}, k}:i_{\operatorname{dep},k}]$.

Let us now state how to set its transmission policy for user $k$ at decision time $i$. 
For the proposed multichannel access, user $k$ sends at most $\min(K[i],N)$ packets to the AP during the time slots $[T[i]:T[i]+K[i]-1]$ based on its action $\mathbf{a}_k[i]$.
To explain how to set  $\mathbf{a}_k[i]$ at decision time $i$ (at the beginning of time slot $t=T[i]$), let $\mathcal{A}[i]\subset [0:N]^{K[i]}$ be the collection of all ordered sets such that $\sum_{j=1}^{K[i]}\mathbf{1}([\mathbf{a}]_j\neq 0)\leq \min(K[i],N)$ for all $\mathbf{a}\in [0:N]^{K[i]}$. Then the action of user $k$ at decision time $i$ (at the beginning of time slot $t=T[i]$) is given by
\begin{align} \label{eq:action_pf}
\mathbf{a}_{k}[i]\in\mathcal{A}[i].
\end{align}
The detailed procedure of choosing $\mathbf{a}_{k}[i]$ will be given in Section \ref{subsec:q_learning}.

\begin{example}
Consider the case where $K[i]=3$ and $N=2$. Then, $\mathcal{A}[i]$ consists of $(0,0,0)$, $(0,0,1)$, $(0,0,2)$, $(0,1,0)$,  $(0,1,1)$,  $(0,1,2)$, $(0,2,0)$, $(0,2,1)$, $(0,2,2)$, $(1,0,0)$, $(1,0,1)$,  $(1,0,2)$,  $(1,1,0)$,  $(1,2,0)$, $(2,0,0)$, $(2,0,1)$, $(2,0,2)$, $(2,1,0)$, and $(2,2,0)$.
For $K[i]=2$ and $N=3$, $\mathcal{A}[i]$ consists of $(0,0)$, $(0,1)$,  $(0,2)$, $(0,3)$, $(1,0)$, $(1,1)$, $(1,2)$, $(1,3)$, $(2,0)$, $(2,1)$, $(2,2)$, $(2,3)$, $(3,0)$, $(3,1)$, $(3,2)$, and $(3,3)$.
\hfill$\lozenge$ 
\end{example}

From the selected $\mathbf{a}_{k}[i]$, user $k$ participates in multichannel access during the time slots $[T[i]:T[i]+K[i]-1]$. Specifically, for $j\in[1:K[i]]$, user $k$ sends its packet via RB $[\mathbf{a}_{k}[i]]_j$ at time slot $t=T[i]+j-1$ if $[\mathbf{a}_{k}[i]]_j\neq 0$ (single packet transmission) and does not send any packet at time slot $t=T[i]+j-1$ if $[\mathbf{a}_{k}[i]]_j=0$ (no transmission). That is, 
\begin{align}
\eta_k(T[i]+j-1)=[\mathbf{a}_{k}[i]]_j
\end{align}
for $j\in[1:K[i]]$, from \eqref{eq:eta_k}.

After the transmission of decision time $i$, i.e., time slots $[T[i]:T[i]+K[i]-1]$, the set of feedback messages $\{b^{[n]}(t)\}_{t\in[T[i]:T[i]+K[i]-1],n\in[1:N]}$ and the set of  instantaneous throughputs $\{\gamma_k(t)\}_{t\in[T[i]:T[i]+K[i]-1]}$ are given from \eqref{eq:ack_nak} and \eqref{eq:ins_th}, respectively. 
The reward of user $k$ at decision time $i$ (at the end of time slot $t=T[i]+K[i]-1]$) is then set as 
\begin{align} \label{eq:reward_pf}
r_k[i]=\sum_{t=T[i]}^{T[i]+K[i]-1}r_k(t),
\end{align}
where
\begin{align} \label{eq:3step_reward}
r_k(t)=\begin{cases}
1 &\mbox{ if } b^{[\eta_k(t)]}(t)=\operatorname{ACK},\\
0 &\mbox{ if } \eta_k(t)=0,\\
-1 &\mbox{ otherwise.}\\
\end{cases}
\end{align}
 

Lastly, we define the state of user $k$ at decision time $i$ (at the beginning of time slot $t=T[i]$)  as
\begin{align} \label{eq:state_pf}
\mathbf{s}_k[i]=\left(\mathbf{a}_k[i-1], \mathbf{r}_k[i-1]\right),
\end{align}
where $\mathbf{r}_k[i]=\big[r_k(T[i]),\cdots, r_k(T[i]+K[i]-1)\big]$.

\subsection{Q-Learning for Multichannel Access} \label{subsec:q_learning}
From the previously defined action, reward, and state functions, we now describe our proposed online Q-learning multichannel access framework. In the following, we briefly introduce a widely adopted Q-learning method and the corresponding deep Q-learning network (DQN).  
Then we explain main technical challenges for adopting such method and state the proposed approach in order to handle those technical issues. 

\subsubsection{Preliminaries}

Q-learning is an empirical value iteration method, that aims on finding an optimal policy to maximize a long-term expected accumulated reward, also known as the `Q-value' ${Q_\pi }\left( {s,a} \right)$, which is a function of a state--action pair $\left( {s,a} \right)$ under policy $\pi$. 
One of the famous model-free RL algorithms is to update the Q-value using the following equation  \cite{49}:
\begin{equation}\label{Q_update}
Q\left( {s,a} \right) \leftarrow \left( {1 - \rho } \right)Q\left( {s,a} \right) + \rho \left[ {r\left( {s,a} \right) + \tau \mathop {\max }\limits_{a' \in \mathcal{A}} Q\left( {s',a'} \right)} \right],
\end{equation}
where $\rho>0$ is the learning rate and $\tau  \in \left[ {0,1} \right]$ is the discount factor that determines the impact of future rewards.

Note that the performance of a Q-learning algorithm depends on the size of the state--action space in a considered system. When the state--action space is small, the agent can easily explore all the state--action pairs in the state--action space and find an optimal action policy. On the other hand, as the size of the state--action space increases, the performance of the Q-learning algorithm is limited because all the state--action pairs in the state--action space may not be explored by the agent.
To overcome the drawback of the Q-learning algorithm described above, DQN can be adopted that exploits deep neural networks (DNN) as a Q-value estimator~\cite{52}. 
More specifically, DQN takes the state--action pair as input and yields the corresponding Q-value as its output. Thus, the optimization of a Q-function $Q\left( {s,a} \right)$ in the Q-learning algorithm is equivalent to the optimization of $\Theta $ in a DNN with $Q\left( {s,a;\Theta } \right)$ where $\Theta$ is the set of weights in the DNN.

\subsubsection{Technical challenges}
If we apply the above Q-learning framework in \eqref{Q_update} to our distributed multichannel access, each user $k$ will take an action at decision time $i$ (at the beginning of time slot $t=T[i]$) such that 
\begin{align} \label{eq:action_q}
\mathbf{a}_k[i]= \mathop{\arg\max}\limits_{\mathbf{a}\in\mathcal{A}[i]} Q_{ (\mathbf{s}_k[i],\mathbf{a} )}[i]
\end{align}
and the Q-value update at decision time $i+1$ (at the beginning of time slot $t=T[i+1]$) will be given as
\begin{align}\label{Q_update_proposed}
Q_{ (\mathbf{s}_k[i],\mathbf{a}_k[i]) }[i+1] = \left( {1 - \rho } \right)Q_{ (\mathbf{s}_k[i],\mathbf{a}_k[i])}[i] + \rho \left[ {r_k[i]+ \tau \mathop {\max }\limits_{\mathbf{a} \in \mathcal{A}[i+1]} Q_{ (\mathbf{s}_k[i+1],\mathbf{a})}[i]} \right],
\end{align}
where $r_k[i]$ is the reward in \eqref{eq:reward_pf} attained by taking $\mathbf{a}_k[i]$ from \eqref{eq:action_q} when $\mathbf{s}_k[i]$ is given.

However, there exist two main technical challenges to be addressed for implementation.
Firstly, it is hard to properly define such update rule in the considered multichannel access problem because the set of active users $\mathcal{K}(T[i])$ is time-varying for each decision time $i$. As a consequence, the sizes of the action and state spaces in \eqref{eq:action_pf} and \eqref{eq:state_pf} are also time-varying, which make implementation of the corresponding Q-value estimator based on DQNs quite challenging. Secondly, the sizes of action and state spaces increase exponentially as either  $N$ or $K[i]$ increases. More specifically, the sizes of the action and state spaces are given by $O(N^{K[i]})$, prohibitively increasing with $N$ and  $K[i]$ compared to the conventional Q-learning based multiaccess strategies operated with space sizes $O(N)$ or $O(N^2)$~\cite{17,28,11}.

As we will verify in Section \ref{sec:DQN}, by assigning separate input nodes for each of the elements in $\mathbf{s}_k[i]$, the number of input nodes in DQNs becomes $O(NK_{\max})$ and therefore we can effectively reduce the complexity of the input layer of DQNs by properly setting the value of $K_{\max}$. To reduce the complexity of the output layer of DQNs, in the following, we propose a vectorized Q-value estimator that provides a fixed and reduced size of the output layer.  
We then will explain how to select an action at each decision time based on such vectorized Q-values.  

\subsubsection{Proposed vectorized Q-learning}
To handle time-varying action and state spaces and provide scalability for a large number of active users and RBs, i.e., for large $N$ and $K[i]$, we first introduce vectorized Q-values with a predetermined size.
Then we present how to set $\mathbf{a}_k[i]$ in a sequential manner based on such vectorized Q-values.
For such purpose, we introduced $K_{\max}$ and defined the decision time and related action, reward, and state functions in Section \ref{subsec:policy}.
Let us now define the Q-value matrix at decision time $i$ as
\begin{align} \label{eq:Q_matrix}
\mathbf{Q}_k[i]\in\mathbb{R}^{(N+1)\times K_{\max}}.
\end{align}
Here, the $(j,l)$th element of $\mathbf{Q}_k[i]$ will be used to represent a Q-value when user $k$ sends a packet via RB $j$ at time slot $T[i]+l-1$.

Algorithm \ref{Algorithm1} states how to  sequentially construct $\mathbf{a}_k[i]$ based on vectorized Q-values in $\mathbf{Q}_k[i]$ at each user $k$ for decision time $i$. After conducting Steps 2 to 7 in Algorithm \ref{Algorithm1}, the maximum number of non-zero elements in $\mathbf{a}_k[i]$ becomes $\min(K[i],N)$, which satisfies the condition for a valid action in \eqref{eq:action_pf}. Suppose that such $\mathbf{a}_k[i]$ is used for the transmission of decision time $i$, which means that each active user in $\mathcal{K}(T[i])$ sends at most $\min(K[i],N)$ packets using $K[i]$ time slots.
Let us first consider the case where $|\mathcal{K}(T[i])|\leq K_{\max}$, i.e., $K[i]=\mathcal{K}(T[i])$. For this case, the maximum per-user throughput and sum throughput are given by $\min\left(1,\frac{N}{|\mathcal{K}(T[i])|}\right)$ and $\min(|\mathcal{K}(T[i])|,N)$, respectively. Therefore, if there is no collision between the users in $\mathcal{K}(T[i])$, the proposed scheme can provide fair throughput between the users in $\mathcal{K}(T[i])$ while maximizing per-user or sum throughputs.

Now consider the case where $|\mathcal{K}(T[i])|> K_{\max}$, i.e., $K[i]=K_{\max}$. As the same manner, the maximum per-user and sum throughputs are given by  $\min\left(1,\frac{N}{K_{\max}}\right)$ and $\min\left(|\mathcal{K}(T[i])|,\frac{N|\mathcal{K}(T[i])|}{K_{\max}}\right)$ respectively for this case.
Then the sum throughput (actually, the sum transmission rate from the users in $\mathcal{K}(T[i]$) can exceed the maximum capacity of the system $N$ and, as a consequence, collision between the users in $\mathcal{K}(T[i])$ is inevitable.    
To prevent such event, we introduce Steps 8 to 10 in Algorithm \ref{Algorithm1} and set each element in $\mathbf{a}_k[i]$ to zero independently with probability $1-\frac{\max(N,K_{\max})}{|\mathcal{K}(T[i])|}$ if $N<|\mathcal{K}(T[i])|$.
From Steps 8 to 10, the maximum expected per-user and sum throughputs are again given by $\min\left(1,\frac{N}{|\mathcal{K}(T[i])|}\right)$ and $\min(|\mathcal{K}(T[i])|,N)$ respectively when  $|\mathcal{K}(T[i])|> K_{\max}$.

\begin{algorithm}[!t] \footnotesize
\caption{Action selection based on vectorized Q-values.}\label{Algorithm1}
\begin{algorithmic}[1]
\State{{\bf Input}: $K[i]$ and $\mathbf{Q}_k[i]$.}
\State{{\bf Initialization}: Define $\mathcal{K}=[1:K[i]]$ and set $\mathbf{a}_k[i]=\mathbf{0}_{K[i]}$.}
\For{$1$ to $\min(K[i],N)$}
\State {Calculate $(a^*, j^*)=\mathop {\arg\max }\limits_{a\in[0:N],j\in\mathcal{K}}[\mathbf{Q}_k[i]]_{(a+1,j)}$.}
\State{Set $[\mathbf{a}_k[i]]_{j^*}=a^*$.}
\State{Update $\mathcal{K}\leftarrow \mathcal{K}\setminus\{j^*\}$.}
\EndFor
\If{$N<|\mathcal{K}(T[i])|$}
\State Set each element in $\mathbf{a}_k[i]$ to zero independently with probability $1-\frac{\max(N,K_{\max})}{|\mathcal{K}(T[i])|}$.
\EndIf
\State{{\bf Output}: $\mathbf{a}_k[i]$.}
 \end{algorithmic}
\end{algorithm}

\begin{algorithm}[!ht] \footnotesize
\caption{Double Q-learning.}\label{Algorithm2}
\begin{algorithmic}[1]
\State{{\bf Initialization}: Define $T_{\operatorname{training}}=0$ and set $T_1$ and $T_2$ as some positive integers.}
\For{decision time $i\in [i_{\operatorname{arr}, k}:i_{\operatorname{dep},k}]$}
\State {The state $\mathbf{s}_k[i]$ is fed into $\mbox{DQN}^{\alpha}_k$ and the corresponding Q-values $\mathbf{Q}^{\alpha}_k[i]$ are obtained as its output.}
\State{Takes an action using the $\epsilon$-greedy strategy. Specifically, choose $\mathbf{a}_k[i]$ from Algorithm \ref{Algorithm1} using $K[i]$ and $\mathbf{Q}^{\alpha}_k[i]$ as its input with probability $1-\epsilon$ or $\mathbf{a}_k[i]$ in $\mathcal{A}[i]$ uniformly at random with probability $\epsilon$.}
\State{After taking $\mathbf{a}_k[i]$,  calculate $\mathbf{r}_k[i]$ and $\mathbf{s}_k[i+1]$ in \eqref{eq:state_pf} from the feedback messages $\mathbf{b}[i]=\big[\mathbf{b}(T[i]),\cdots, \mathbf{b}(T[i]+K[i]-1)\big]$.}
\State{The state $\mathbf{s}_k[i+1]$ is fed into both $\mbox{DQN}^{\alpha}_k$ and $\mbox{DQN}^{\beta}_k$. Denote $\mathbf{Q}^{\alpha}_k[i+1]$ and $\mathbf{Q}^{\beta}_k[i+1]$ as the corresponding outputs of $\mbox{DQN}^{\alpha}_k$ and $\mbox{DQN}^{\beta}_k$, respectively.}
\State{Update a subset of Q-values in $\mathbf{Q}^{\alpha}_k[i]$ obtained in Step 2 as follows:}
\For{$j\in[1:\min(K[i],N)]$}
\State{$[\mathbf{Q}^{\alpha}_k[i]]_{([\mathbf{a}_k[i]]_j+1,j)}\leftarrow[\mathbf{r}_k[i]]_j+\tau[\mathbf{Q}^{\beta}_k[i+1]]_{(a^*+1,j)}$, where $a^*=\mathop {\arg\max }\limits_{a\in[0:N]}[\mathbf{Q}^{\alpha}_k[i+1]]_{(a+1,j)}$.}
\EndFor
\State{Store $\mathbf{s}_k[i]$ and $\mathbf{Q}^{\alpha}_k[i]$ in its buffer memory.}
\State{Update $T_{\operatorname{training}}\leftarrow T_{\operatorname{training}}+1$.}
\If{$T_{\operatorname{training}}/ T_1=0$}
\State{Using data samples in the buffer memory, train $\mbox{DQN}^{\alpha}_k$ by treating $\mbox{DQN}^{\beta}_k$ as a target DQN.}
\EndIf
\If{$T_{\operatorname{training}}/ (T_1T_2)=0$}
\State{Update the weights in $\mbox{DQN}^{\beta}_k$ with the weights in $\mbox{DQN}^{\alpha}_k$.}
\EndIf
\EndFor
 \end{algorithmic}
\end{algorithm}

\section{Deep Q-Network Training}\label{sec:DQN}

In this section, we introduce a DQN with a fixed set of weights $\Theta$ to estimate the vectorized Q-values $\mathbf{Q}_k[i]$ at each decision time $i$ introduced in Section \ref{subsec:q_learning} when the state $\mathbf{s}_k[i]$ is given.  
It is worthwhile to mention that each active user has its own DQN and initiates DQN training based on its local observation in a fully distributed and decentralized manner.

\subsection{DQN Training Algorithm}
We adopt the double Q-learning method to avoid overestimated Q-values calculated from the executed actions~\cite{49}. That is, for the purpose of stabilization, the double-Q learning method utilizes two separate DQNs, one for action execution and the other for Q-value estimation used for training.
In particular, each user $k$ operates two separate DQNs, denoted by $\mbox{DQN}^{\alpha}_k$ and $\mbox{DQN}^{\beta}_k$.
The detailed DQN structure is provided in Section \ref{subsec:DQN}.
The first DQN, $\mbox{DQN}^{\alpha}_k$, will be used for choosing and executing an action, while the second DQN, $\mbox{DQN}^{\beta}_k$, will be used to estimate the corresponding Q-values associated with the selected action. At each decision time, an action is selected based on the Q-values attained from $\mbox{DQN}^{\alpha}_k$. Then the next state is fed into $\mbox{DQN}^{\beta}_k$ and a subset of attained Q-values are used for Q-value update and the updated Q-values are stored in buffer memory.

Algorithm \ref{Algorithm2} states the proposed action selection algorithm and DQN training methodology.
Each user $k$ performs Algorithm \ref{Algorithm2} in a distributed manner during its active time duration, i.e., during the decision times $[i_{\operatorname{arr}, k}:i_{\operatorname{dep},k}]$.

\subsection{DQN Architecture} \label{subsec:DQN}
\begin{figure}[t] \centering
\includegraphics[scale=0.6]{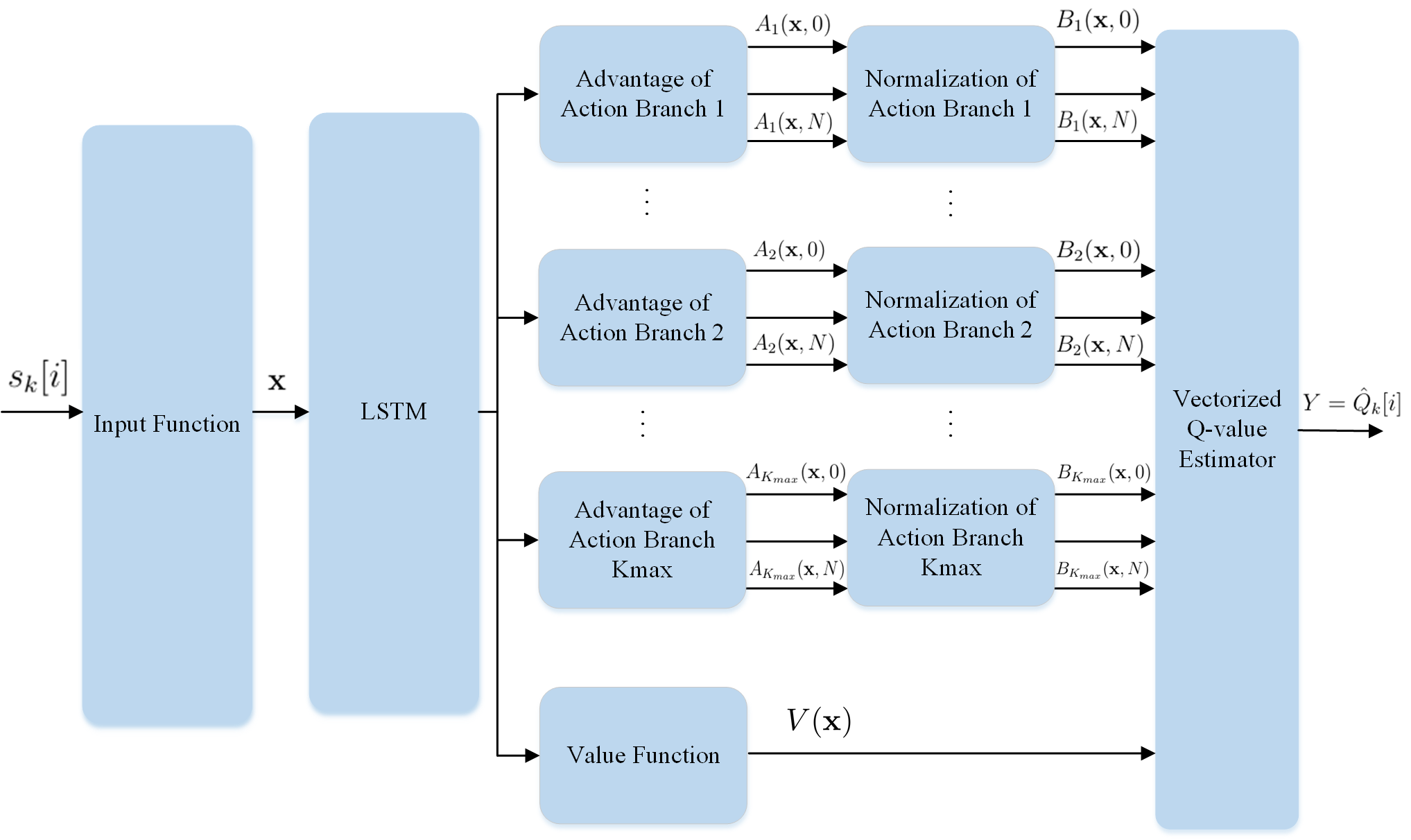}
\caption{Proposed DQN architecture.}\label{fig:DNN_architecture}
\end{figure}

In this subsection, we state the proposed DQN architecture applied for both $\mbox{DQN}^{\alpha}_k$ and $\mbox{DQN}^{\beta}_k$. 
For notational convenience, we omit the superscripts $\alpha$ and $\beta$ in this subsection.
Fig. \ref{fig:DNN_architecture} illustrates the overall DQN architecture used at each user $k$. We adopt a dueling network architecture~\cite{15,11}, which separates the state values and state-dependent action advantages using two independent streams. 
Specifically, denote the value of being in state $s$ by $V(s)$ and the advantage of taking action $a$ in state $s$ by $A(s, a)$.
The Q-value can therefore be expressed as  $Q(s,a)=V(s)+A(s,a)$. Here, the generalized formula for the value function is given by $V(s)=E \big[\sum_{i=1}^{T}\gamma^{i-1}r_{i} \big]$, where $\gamma\in(0,1]$ is the discount factor and $r_i$ is the reward at time $i$. Notice that $V(s)$ is only depends on a policy of an agent. Hence, by introducing two separate estimators, the dueling architecture can learn which states are (or not) good, without having to learn the effect of an executed action for each state. That is, each of two estimators estimates $V(s)$ and $A(s,a)$ separately and then the corresponding Q-value is generated from two estimated values, which can be helpful to improve the accuracy of the estimated Q-value compared to the conventional DQNs.
In order to efficiently estimate vectorized Q-values, which correspond to the elements in \eqref{eq:Q_matrix}, we have modified the original dueling network architecture in order to estimate vectorized Q-values with low computational complexity.
In particular, the proposed DQN architecture generates multiple branches of state and action advantage values and then obtains estimated vectorized Q-values from those multiple branches.

\subsubsection{Input layer} \label{subsubsec:input}
Let $\mathbf{x}\in\{0,1\}^{(N+1)K_{\max}}\times \{-1,0,1\}^{K_{\max}}$ be the input layer of $\mbox{DQN}_k$.
Firstly, we explain how to construct $\mathbf{x}$ from the state $\mathbf{s}_k[i]$ at decision time $i$.
At decision time $i$, $\mathbf{s}_k[i]$ in \eqref{eq:state_pf} will be applied to $\mbox{DQN}_k$.
Recall that $\mathbf{a}_k[i]\in [0:N]^{K[i]}$ and $\mathbf{r}_k[i]\in\{-1,0,1\}^{K[i]}$.
Then by applying one-hot encoding for each element in $\mathbf{a}_k[i]$, we have $\mathbf{a}'_k[i]\in \{0,1\}^{(N+1)K[i]}$.
Therefore, for $K[i]=K_{\max}$, the input function is given by $\mathbf{x}=(\mathbf{a}'_k[i-1], \mathbf{r}_k[i-1])\in\{0,1\}^{(N+1)K_{\max}}\times \{-1,0,1\}^{K_{\max}}$ at decision time $i$.
For $K[i]<K_{\max}$, the input function is given by $\mathbf{x}=(\mathbf{a}'_k[i-1], \mathbf{0}_{(N+1)(K_{\max}-|\mathcal{K}(T[i-1])|)}, \mathbf{r}_k[i-1], \mathbf{0}_{(K_{\max}-|\mathcal{K}(T[i-1])|)})$ at decision time $i$. That is, zero-padding is applied for this case.

\subsubsection{Long short-term memory (LSTM) layer} 
After the input layer,  we introduce an LSTM layer, which maintains internal states and aggregates observations over time slots. It helps the DQN in estimating the correct state from the related historical information. This layer is responsible of how to aggregate the observations over time slots.
The output of the LSTM layer will be applied to the input of both the action advantage estimator and value estimator as seen in Fig. \ref{fig:DNN_architecture}.

\subsubsection{Value and advantage layers} 
As mentioned previously, we incorporate a dueling network architecture, which separates the state values and state-dependent action advantages using two separate streams. Hence, it is desirable to estimate the average Q-value of the state $\mathbf{s}_k[i]$ which is called the value of the state $V(\mathbf{s}_k[i])$ measured independently from the advantage of an executed action $\mathbf{a}_k[i]$. 
Previously, we explained how to construct the input layer $\mathbf{x}$ having a fixed size from $\mathbf{s}_k[i]$ having different sizes for each decision time $i$.
Instead of defining $V(\mathbf{s}_k[i])$ based on $\mathbf{s}_k[i]$, we introduce $V(\mathbf{x})$ for the proposed DQN architecture, which is a function of the input layer $\mathbf{x}$.

However, it is still not straightforward how to properly define the advantage of an executed action since the size of the action space $\mathcal{A}[i]$ is time-varying for each decision time $i$.
To handle this technical issue, we adapt a branching dueling Q-network architecture~\cite{50} and measure the advantage of an executed action separately by each time slot. In particular, we consider action branches $[1:K_{\max}]$, i.e., time slots $[1:K_{\max}]$. 
For each action branch $j\in[1:K_{\max}]$, let $A_j(\mathbf{x},a_j)$ be the sub-action advantage for the $j$th sub-action $a_j\in[0:N]$ when the input $\mathbf{x}$ is applied.
As seen in Fig. \ref{fig:DNN_architecture}, $A_j(\mathbf{x},0)$ to $A_j(\mathbf{x},N)$ are the set of outputs of the advantage estimator of action branch $j$. 
This set of outputs is then normalized to construct $B_j(\mathbf{x},a_j)=A_j(\mathbf{x},a_j)-\frac{1}{N+1}\sum_{a'_j \in [0:N]}A_j(\mathbf{s},a'_j)$ for $a_j\in[0:N]$, which correspond to the outputs of the normalization function of action branch $j$.

\subsubsection{Output layer} Let $\mathbf{Y}\in\mathbb{R}^{(N+1)K_{\max}}$ be the output of $\mbox{DQN}_k$, which will be used to estimate $\mathbf{Q}_k[i]$ in \eqref{eq:Q_matrix}. From $V(\mathbf{x})$ and $A_j(\mathbf{x}, a_j)$, the corresponding Q-value for sub-action $a_j\in[0:N]$ at the input $\mathbf{x}$ (when $\mathbf{s}_k[i]$ applied at decision time $i$) is expressed as a combination of the common input (state) value $V(\mathbf{x})$ and the normalized sub-action advantage $B_j(\mathbf{x}, a_j)$ using the following equation: 
\begin{align} \label{eq:Q_j}
Q_j(\mathbf{x}, a_j)=V(\mathbf{x})+B_j(\mathbf{x},a_j).
\end{align}
Then we set $[\mathbf{Y}]_{(l+1,j)}=Q_j(\mathbf{x}, a_j=l)$ for all $j\in[1:K_{\max}]$ and $l\in[0:N]$. 
Finally, we use such $\mathbf{Y}$ as the estimated $\mathbf{Q}_k[i]$ at decision time $i$.
In summary, at decision time $i$, each user $k$ applies $\mathbf{s}_k[i]$ to $\mbox{DQN}_k$ and attains the estimated $\mathbf{Q}_k[i]$.

\section{Numerical Evaluation} \label{sec:result}
In this section, we evaluate the performance of our proposed algorithm.
The architecture of DQN which we used in the algorithm consists of
an LSTM layer with 300 neurons.
A fully connected neural network consisting of 50 neurons at the first layer and 1 neuron at the second layer is  used for calculating the value of state, i.e., $V(\mathbf{x})$ and another fully connect network consisting of $K_{\max}$ layers with $N+1$ neurons in each layer is used for calculating the advantages of actions, i.e., $A_j(\mathbf{s},a_j)$.
Finally, the output layer consists of $K_{\max}$ neurons, each of which corresponds to the Q-value of a certain action.
The deep RL algorithm uses the epsilon greedy strategy with $\epsilon$ initially set to $0.1$ in order to maintain a balance between exploitation and exploration, and it is decreased after every training to make the algorithm more greedy. The values of several important hyperparameters of the DQN are provided in table \ref{tab:table1}.
\begin{table}[H]
  \begin{center}
    \caption{List of DQN hyperparameters.}
    \label{tab:table1}
	\scalebox{0.85}{
    \begin{tabular}{l|c} 
     \hline 
	  \multicolumn{1}{c|}{Hyperparameter} & Value\\
      \hline\hline
      \multicolumn{1}{c|}{Learning rate ($\rho$)} & 0.01\\
		\hline
      \multicolumn{1}{c|}{Activation function} & Relu\\
		\hline
		\multicolumn{1}{c|}{Optimizer} & Adam\\
		\hline
		\multicolumn{1}{c|}{Loss function} & Mean squared error\\
       \hline
      \multicolumn{1}{c|}{Discount factor ($\tau$)}  & 0.95\\
		\hline
		\multicolumn{1}{c|}{$\epsilon$ for $\epsilon$-greedy} & 0.1\\
		\hline
		\multicolumn{1}{c|}{Minibatch size} & 40 \\
		\hline
    \end{tabular}}
  \end{center}
\end{table}	 

\subsection{Fixed User Environment}
\begin{figure}[H]
\centering

     \begin{subfigure}[b]{0.32\textwidth}
         \includegraphics[width=1.05\textwidth, scale=3, height=1.8in]{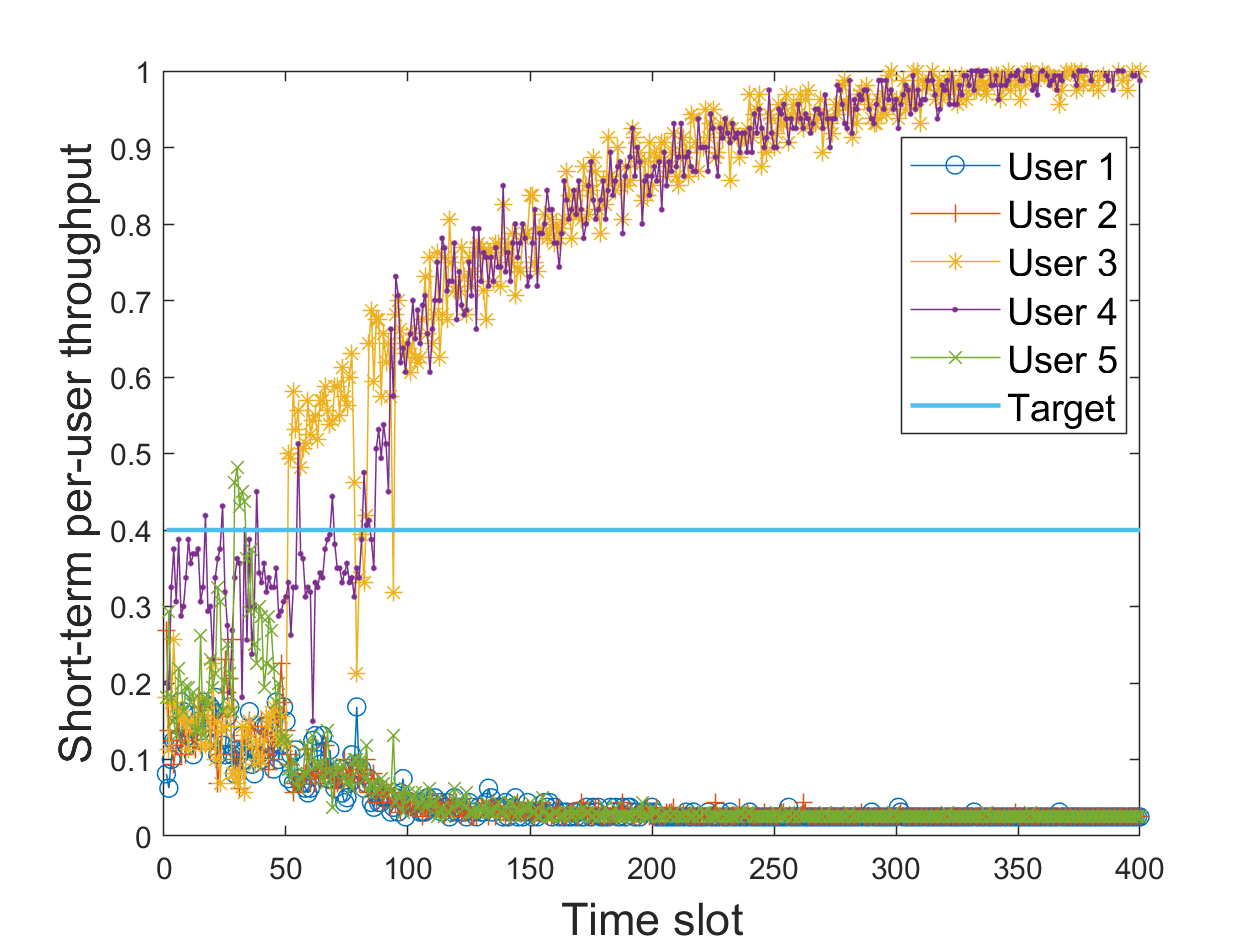}
         \caption{Conventional RL}
     \end{subfigure}
\hfill
\begin{subfigure}[b]{0.32\textwidth}
         \includegraphics[width=1.05\textwidth, scale=3, height=1.8in]{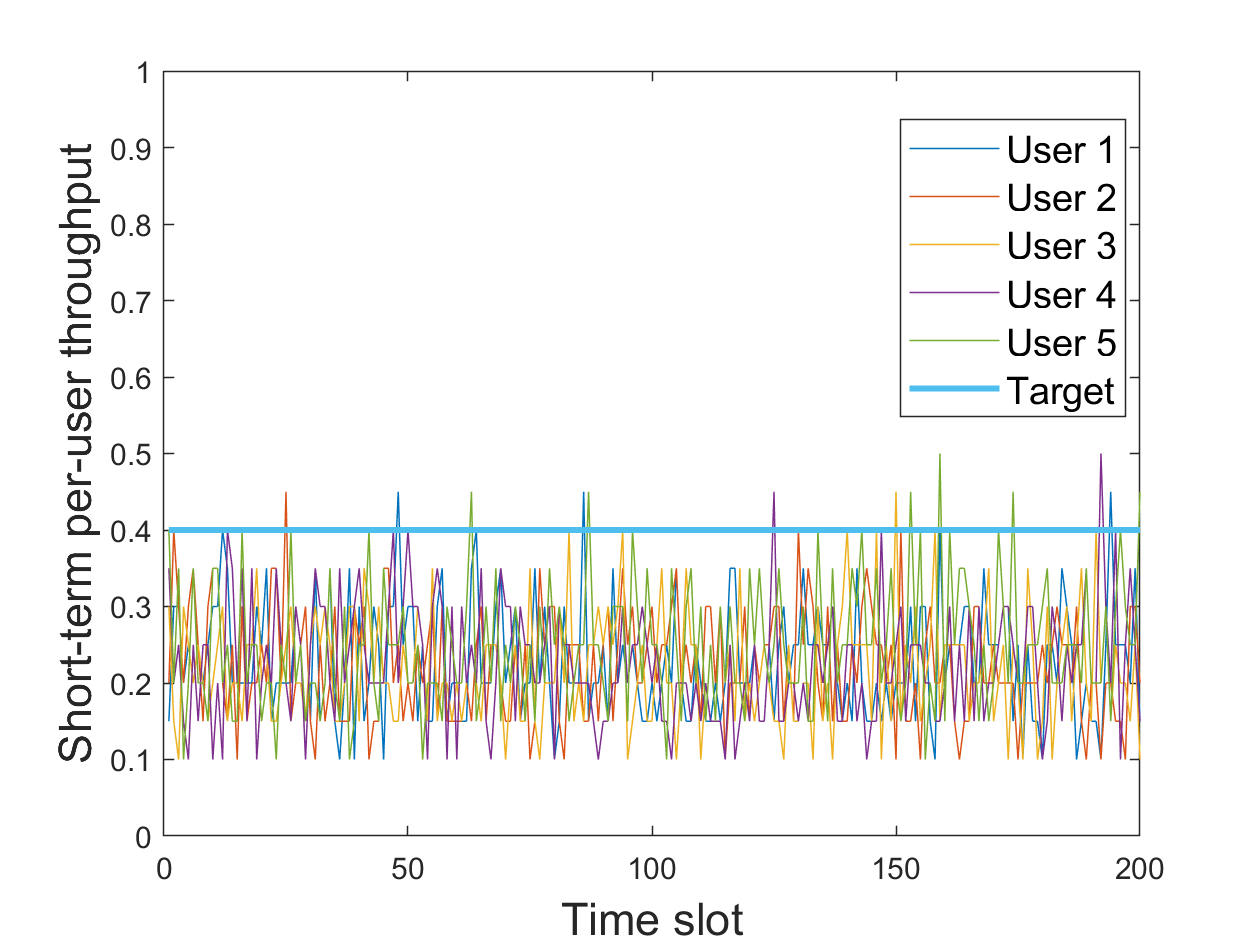}
         \caption{Conventional RL with PF}
     \end{subfigure}
\hfill
     \begin{subfigure}[b]{0.32\textwidth}
         \includegraphics[width=1.05\textwidth, scale=3, height=1.8in]{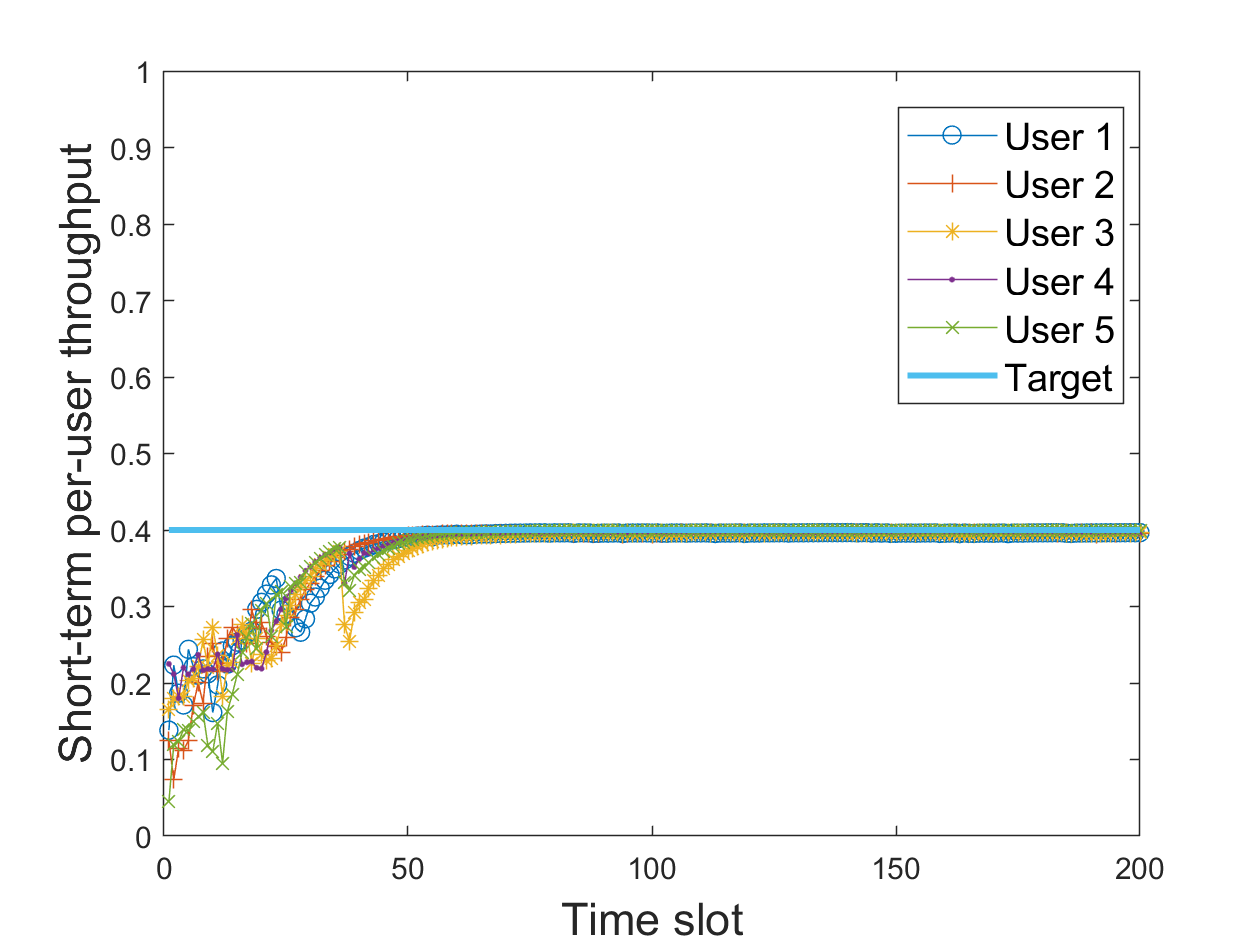}
         \caption{Proposed scheme                                   }
     \end{subfigure}
        \caption{$\Gamma_k(t)$ with respect to time slot $t$ for $k\in[1:K]$.}
        \label{fig:gamma_k}
\end{figure}

Let us first consider the case where the number of users is given as $K$ and fixed over the entire simulation epoch, i.e., $\mathcal{K}(t)=[1:K]$ for all $t$.
Then, from \eqref{eq:th_target} and \eqref{eq:th_target_ave}, $\Gamma_{\operatorname{target},k}(t)=\min\left(1,\frac{N}{K}\right)$ for this case regardless of the average window size $T_{w}$.
Fig. \ref{fig:gamma_k} (a) plots the short-term achievable throughput $\Gamma_k(t)$ in \eqref{eq:th_ave} for $K=5$ and $N=2$, attained by applying the conventional RL approach~\cite{17,28} in which each agent adjusts its transmission policy separately at each time slot, where we set $T_w=20$ in the figure.
The conventional RL approach can achieve the average sum throughput close to its upper bound $\min(N,K)$, improving the sum throughput performance significantly compared to the ALOHA cases without RL~\cite{43,44,45,46}. 


%
%

However, fairness between users is hard to guarantee as seen in the figure because the throughput of each user is biased from the initial channel access trials between users.
To handle such limitation, PF has been introduced in \cite{11} for taking actions in each distributed agent. 
Fig. \ref{fig:gamma_k} (b) plots $\Gamma_k(t)$ of the conventional RL with PF scheme for $K=5$ and $N=2$, where $T_w=20$.
As seen in the figure,  although it provides enhanced fairness between users, the achievable per-user throughput is decreased to $0.23$, providing the sum throughput of $1.15$.
Fig. \ref{fig:gamma_k} (c) plots $\Gamma_k(t)$ of the proposed scheme  for $K=5$ and $N=2$, where we set $K_{\max}=K$ in simulation and $T_w=20$. As seen in the figure, the proposed scheme is able to provide both improved sum throughput and fairness.  

Tables \ref{tab:table 2} presents the short-term average throughput loss $\frac{1}{K}\sum_{k=1}^{K} \Delta_k$ and the long-term average sum throughput $\frac{1}{T}\sum_{k=1}^K\sum_{t=1}^T\gamma_k(t)$ of the conventional RL, the conventional RL with PF, and  the proposed scheme respectively for various $K$ and $N$.\footnote{For the fixed user case, the weighted average in \eqref{eq:min_optimization} becomes the ordinary average since $T_k$ is the same for all $k\in[1:K]$.} As seen in Fig. \ref{fig:gamma_k},  throughputs of each user converge to certain values as the number of time slots increases. Hence, throughput losses with and without the initial learning period will be eventually the same as the simulation time increases.
Numerical results demonstrate that the proposed scheme provides an improved short-term fairness compared to the conventional RL and the conventional RL with PF. It also improves the long-term average sum throughput at the same time.

\begin{table}\caption{Throughput comparison.}\label{tab:table 2}
\begin{center}
\begin{subtable}[h]{0.45\textwidth}
\begin{center}
\scalebox{0.80}{
\begin{tabular}{c|c|c|c|c}
\hline
\multicolumn{1}{c|}{(K , N)} & \multicolumn{3}{c|}{\textbf{$\frac{1}{K}\sum_{k=1}^{K}\Delta_k$}} & \multicolumn{1}{c}{\textbf{$\frac{1}{T}\sum_{k=1}^{K}\sum_{t=1}^{T}\gamma_k(t)$}} \\
\hline
\hline
\multirow[|c]{2}{*}{(5 , 2)} & {$T_w=5$} & {$T_w=10$} & {$T_w=20$} & \multirow[c]{2}{*}{$1.61$}\\
\cline{2-4}
&  $0.220$ &  $0.219$ &$0.218$\\
\hline
\multirow[|c]{2}{*}{(10 , 2)} & {$T_w=5$} & {$T_w=10$} & {$T_w=20$} & \multirow[c]{2}{*}{$1.46$}\\
\cline{2-4}
& $0.147$ & $0.145$ &$0.144$\\
\hline
\multirow[|c]{2}{*}{(10 , 4)} & {$T_w=5$} & {$T_w=10$} & {$T_w=20$} & \multirow[c]{2}{*}{$2.91$}\\
\cline{2-4}
& $0.239$ & $0.236$ &$0.234$\\
\hline
\end{tabular}}
\end{center}
\caption{Conventional RL}
\end{subtable}
\hfill

\vspace*{5mm}

\begin{subtable}[h]{0.45\textwidth}
\begin{center}
\scalebox{0.80}{
\begin{tabular}{c|c|c|c|c}
\hline
\multicolumn{1}{c|}{(K , N)} & \multicolumn{3}{c|}{\textbf{$\frac{1}{K}\sum_{k=1}^{K}\Delta_k$}} & \multicolumn{1}{c}{\textbf{$\frac{1}{T}\sum_{k=1}^{K}\sum_{t=1}^{T}\gamma_k(t)$}} \\
\hline
\hline
\multirow[|c]{2}{*}{(5 , 2)} & {$T_w=5$} & {$T_w=10$} & {$T_w=20$} & \multirow[c]{2}{*}{$1.15$}\\
\cline{2-4}
&  $0.169$ &  $0.163$ &$0.151$\\
\hline
\multirow[|c]{2}{*}{(10 , 2)} & {$T_w=5$} & {$T_w=10$} & {$T_w=20$} & \multirow[c]{2}{*}{$0.53$}\\
\cline{2-4}
& $0.147$ & $0.146$ &$0.144$\\
\hline
\multirow[|c]{2}{*}{(10 , 4)} & {$T_w=5$} & {$T_w=10$} & {$T_w=20$} & \multirow[c]{2}{*}{$1.57$}\\
\cline{2-4}
& $0.242$ & $0.0.240$ &$0.236$\\
\hline
\end{tabular}}
\end{center}
\caption{Conventional RL with PF}
\end{subtable}

\hfill
\vspace*{5mm}

\begin{subtable}[h]{0.55\textwidth}
\begin{center}
\scalebox{0.80}{
\begin{tabular}{c|c|c|c|c}
\hline
\multicolumn{1}{c|}{(K , N)} & \multicolumn{3}{c|}{\textbf{$\frac{1}{K}\sum_{k=1}^{K}\Delta_k$}} & \multicolumn{1}{c}{\textbf{$\frac{1}{T}\sum_{k=1}^{K}\sum_{t=1}^{T}\gamma_k(t)$}} \\
\hline
\hline
\multirow[|c]{2}{*}{(5 , 2)} & {$T_w=5$} & {$T_w=10$} & {$T_w=20$} & \multirow[c]{2}{*}{$1.94$}\\
\cline{2-4}
&  $0.050$ &  $0.032$ &$0.022$\\
\hline
\multirow[|c]{2}{*}{(10 , 2)} & {$T_w=5$} & {$T_w=10$} & {$T_w=20$} & \multirow[c]{2}{*}{$1.91$}\\
\cline{2-4}
& $0.033$ & $0.021$ &$0.014$\\
\hline
\multirow[|c]{2}{*}{(10 , 4)} & {$T_w=5$} & {$T_w=10$} & {$T_w=20$} & \multirow[c]{2}{*}{$3.89$}\\
\cline{2-4}
& $0.049$ & $0.031$ &$0.021$\\
\hline
\end{tabular}}
\end{center}
\caption{Proposed scheme }
\end{subtable}
\end{center}

\end{table}

\begin{figure}[H]
\center
         \includegraphics[width=3in]{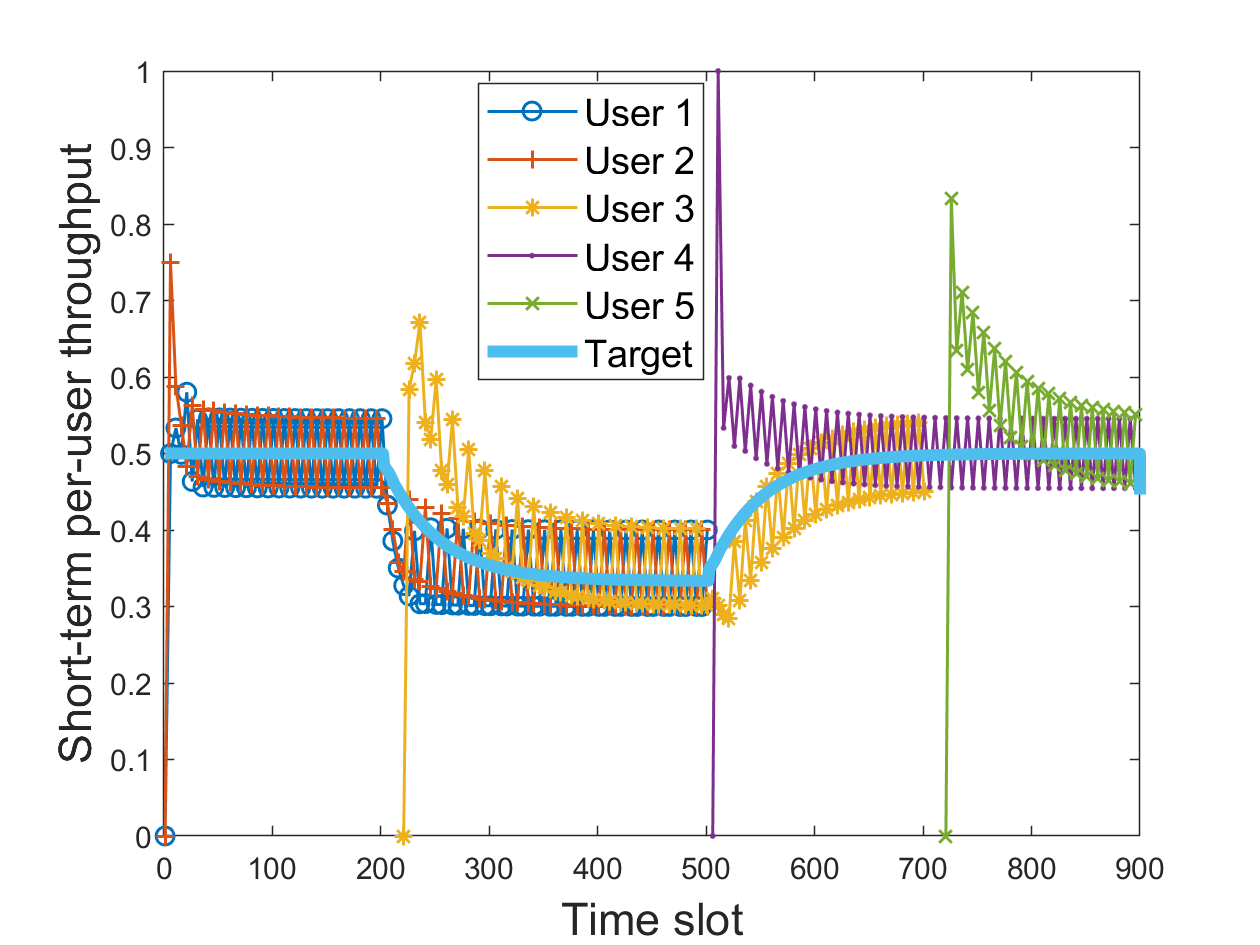}
         \caption{Short-term per-user throughputs with respect to time slot $t$.}
         \label{fig:short_term_th}
\end{figure}

\subsection{Dynamic User Environment}
Let us now consider the dynamic user environment in which each user arrives randomly in the system and be activated for a certain period of time slots and then disappear from the system.
It would be worthwhile to emphasize that  the primary aim of this paper is to provide fair per-user throughputs between active users in real-time by adjusting multi-channel access policies in a distributed manner from online RL agents. 
For such purpose, in simulation, we assume that new users arrive according to Poisson arrivals with rate $\lambda$ at each time slot. The activation time of each user is assumed to follow a uniform distribution between $[T_{\min}:T_{\max}]$. 
Note that for the average number of active users is then given by  $\frac{T_{\max}-T_{\min}+1}{2}\lambda$, i.e., $E(|\mathcal{K}(t)|)=\frac{T_{\max}-T_{\min}+1}{2}\lambda$ for this case.
 
To demonstrate the dynamic adjustment of the proposed multi-channel access, Fig. \ref{fig:short_term_th} plots an example episode for short-term per-user throughputs of each active user with respect to time slots, where we set $T_w=20$ in the figure. For comparison, we also plot the short-term average target per-user throughput. As seen in the figure, each user adjusts its transmission policy in real-time to provide fair resource sharing between active users.

\begin{figure}[H]
     \begin{subfigure}{0.5\textwidth}
			\centering
         \includegraphics[width=2.4in]{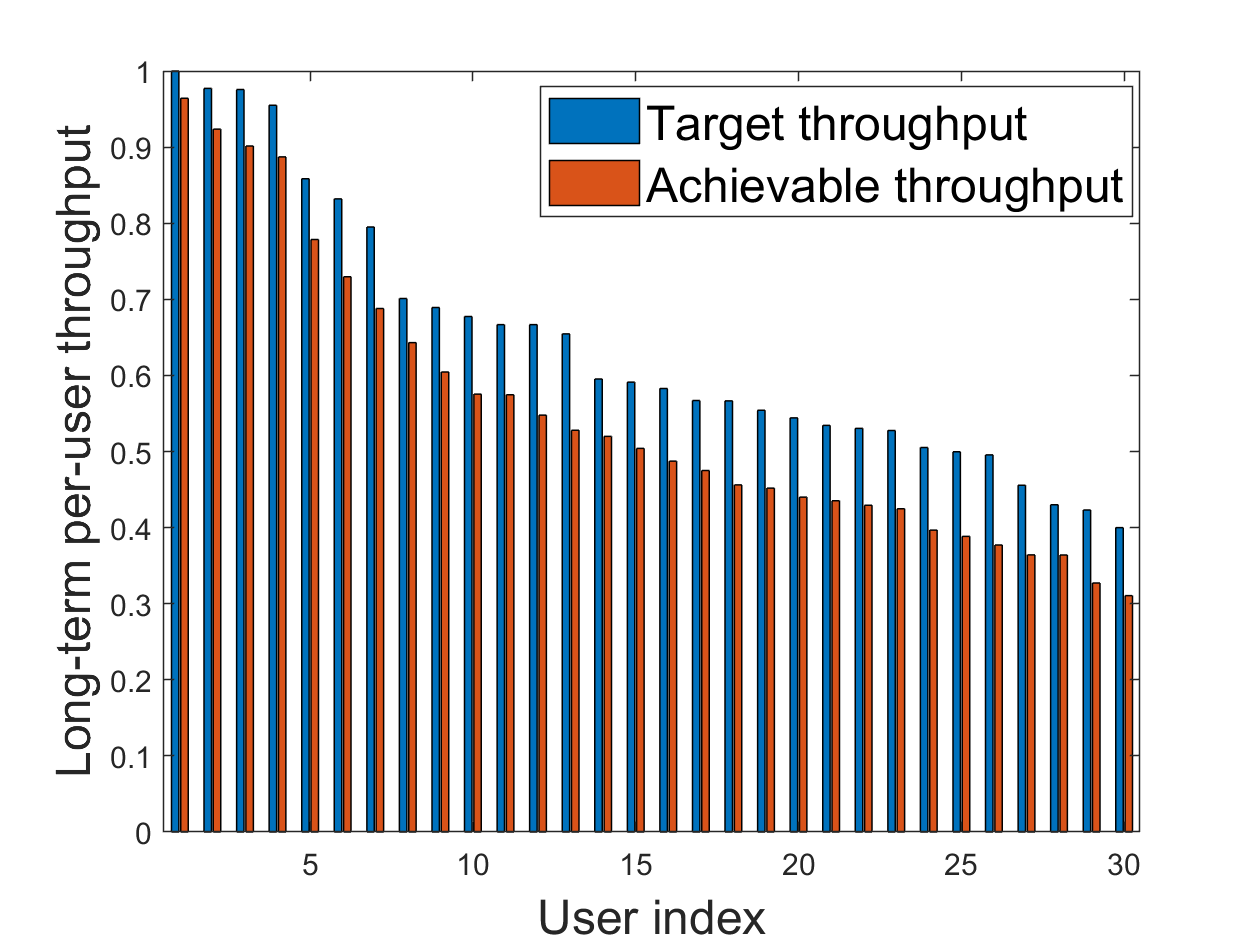}
         \caption{$\lambda=0.02$, $N=2$}
     \end{subfigure}
\begin{subfigure}{0.5\textwidth}
			\centering
         \includegraphics[width=2.4in]{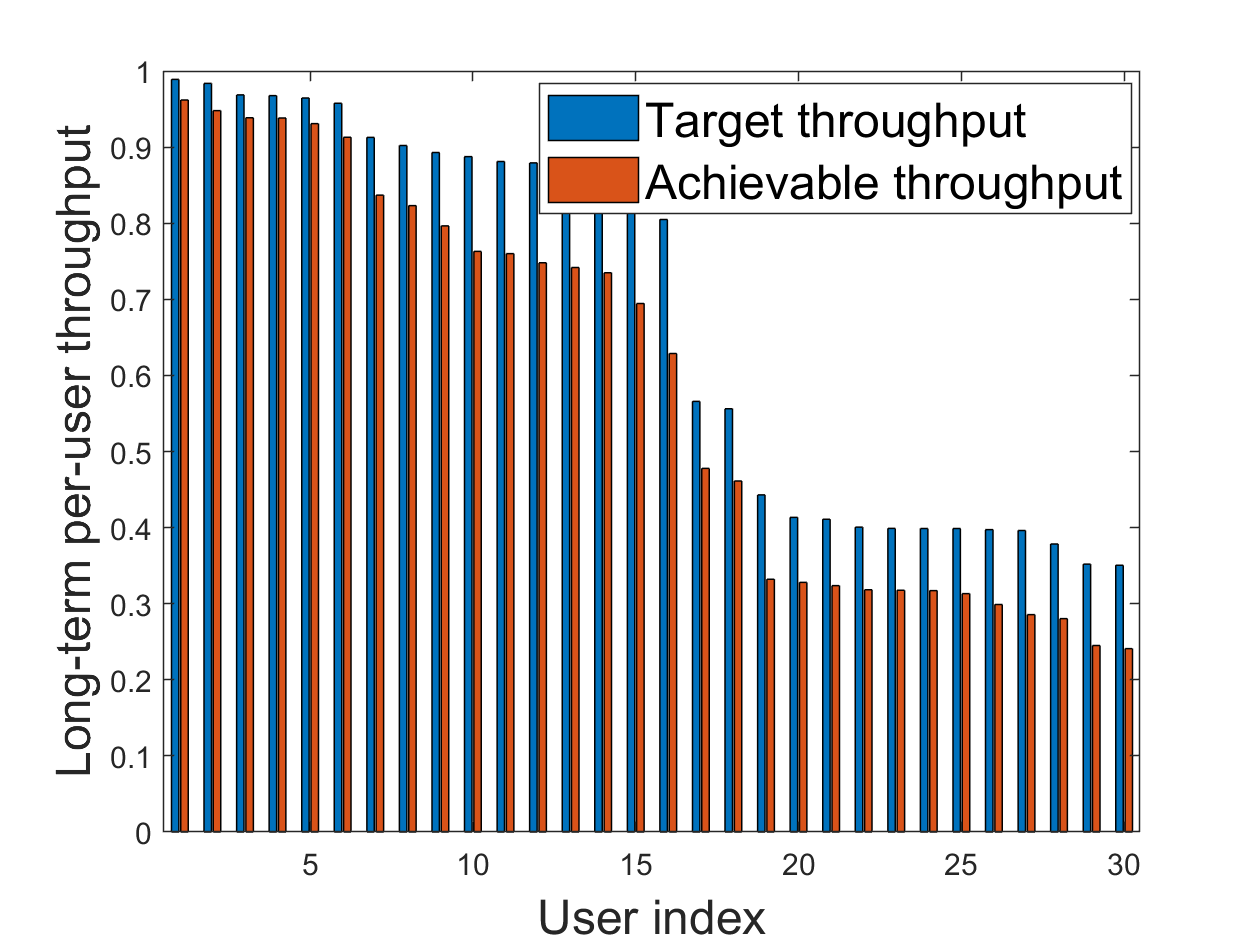}
         \caption{$\lambda=0.02$, $N=4$}
     \end{subfigure}
     \begin{subfigure}{0.5\textwidth}
		\centering
         \includegraphics[width=2.4in]{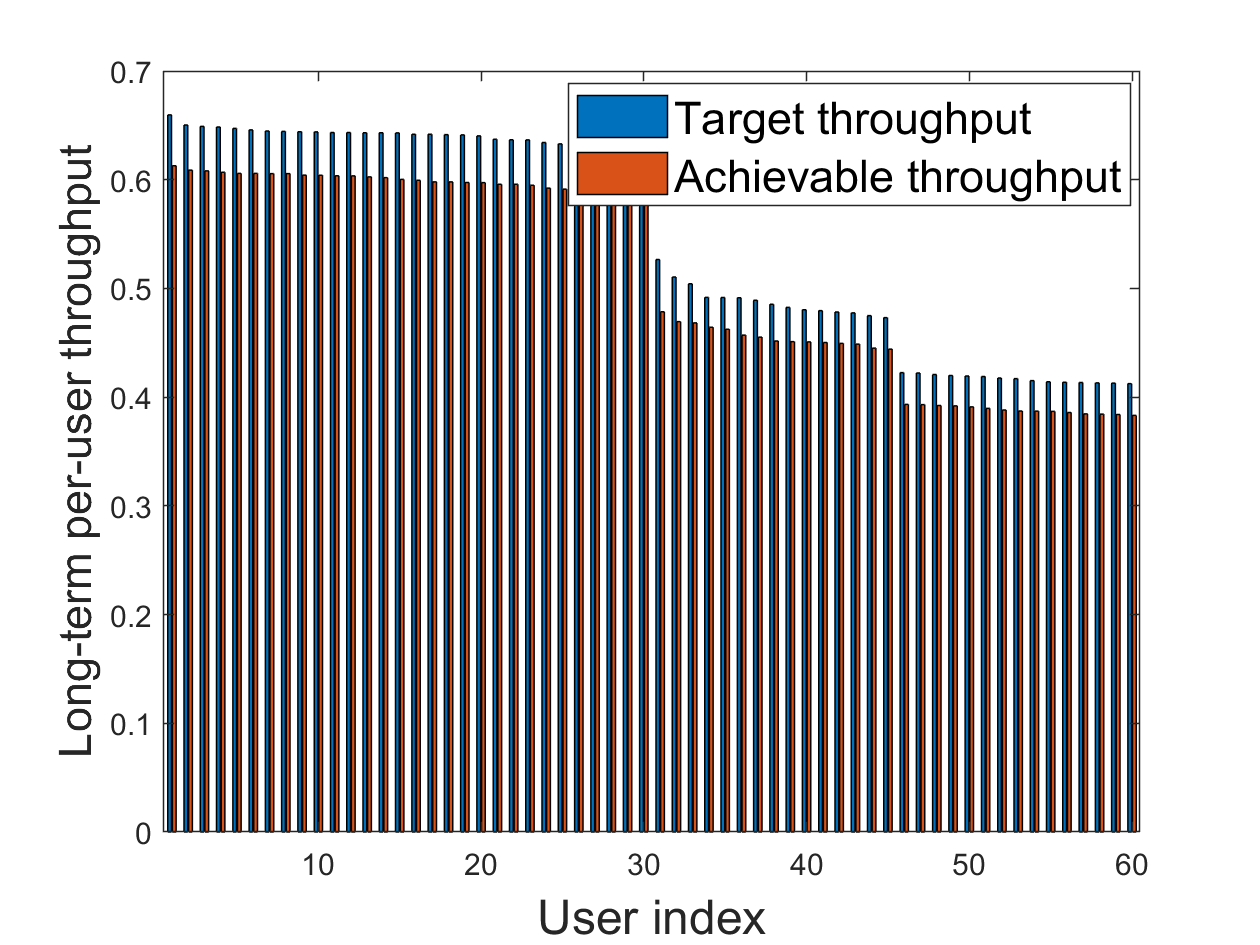}
         \caption{$\lambda=0.2$, $N=20$}
         \label{fig:five over x}
     \end{subfigure}
     \begin{subfigure}{0.5\textwidth}
		 \centering
         \includegraphics[width=2.4in]{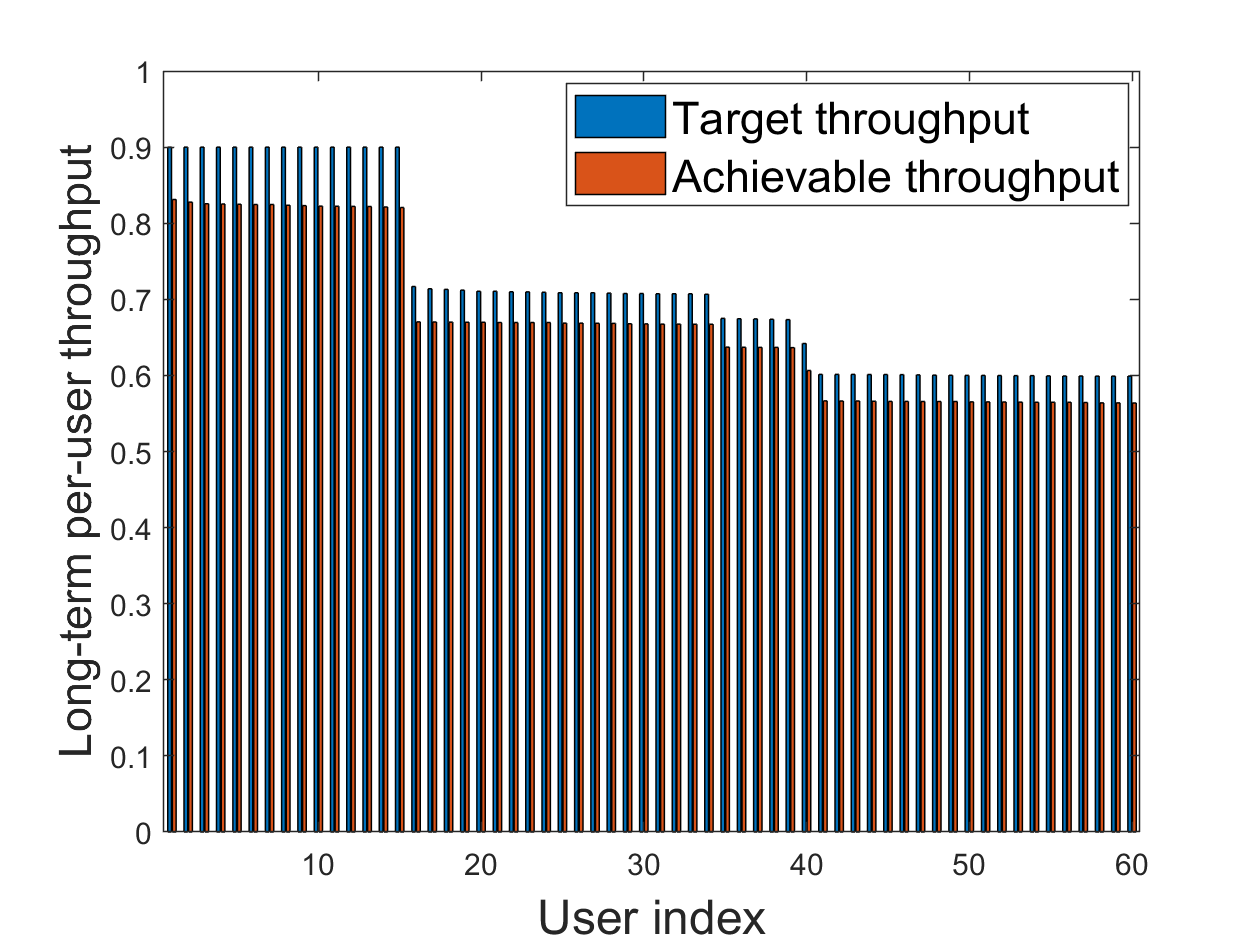}
         \caption{$\lambda=0.2$, $N=50$}
         \label{fig:five over x}
     \end{subfigure}
        \caption{Long-term per-user throughputs.}
        \label{fig:three graphs}
\end{figure}


\begin{figure}[H] 
			\centering
         \includegraphics[width=3in]{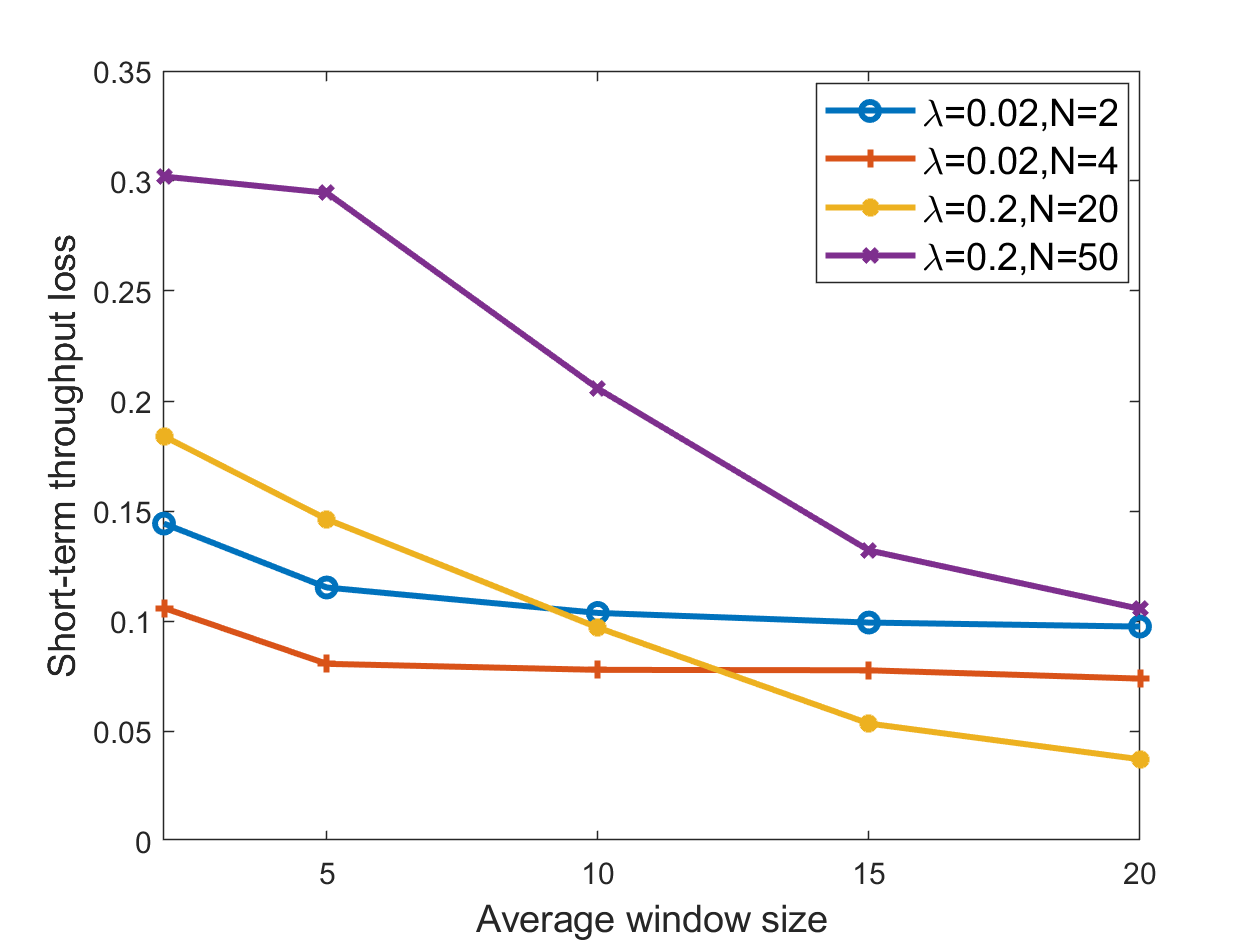}
         \caption{$\Delta_k$ with respect to the average window size $T_w$.}
         \label{fig:convention}
\end{figure}

Fig. \ref{fig:three graphs} plots long-term average per-user throughputs $\frac{1}{T_k}\sum_{t=t_{\operatorname{arr}, k}}^{t_{\operatorname{dep},k}}\gamma_k(t)$ attained by the proposed scheme when $\lambda=0.02$ and $N=2$ (Fig. \ref{fig:three graphs} (a)) , $\lambda=0.02$ and $N=4$ (Fig. \ref{fig:three graphs} (b)), $\lambda=0.2$ and $N=20$ (Fig. \ref{fig:three graphs} (c)), and  $\lambda=0.2$ and $N=50$ (Fig. \ref{fig:three graphs} (d)). In simulation, we set $T_{\min}$ and $T_{\max}$ such that the number of decision times for each user becomes approximately $100$ to $200$ for the first two cases and $300$ to $400$ for the other two cases.
For such setting, $E(|\mathcal{K}(t)|)$ is approximately given by $3$ for the first two cases and  $40$ for the other two cases.
Considering these expected numbers of active users, we set $K_{\max}=5$ for the first two cases and $K_{\max}=50$ for the other two cases.
For comparison, we also plot long-term average target per-user throughputs  $\frac{1}{T_k}\sum_{t=t_{\operatorname{arr}, k}}^{t_{\operatorname{dep},k}}\gamma_{\operatorname{target}}(t)$, which is a performance upper bound for user $k$.
Note that because the number of active users is different for each user, the long-term average per-user throughput and its upper bound are also different for each user.
For convenience, we plot the per-user throughputs of the first $30$ active users in Fig. \ref{fig:three graphs} (a) and (b) and $60$ active users in Fig. \ref{fig:three graphs} (c) and (d), respectively. 
As seen in the figures, the proposed scheme is able to provide real-time fair per-user throughputs for each active user even for such dynamic environment.
More specifically, Fig. \ref{fig:convention} plots the short-term average throughput loss $\Delta_k$ with respect to the average window size $T_w$.
The proposed scheme guarantees $\Delta_k$ less than $0.15$ when $T_w\geq 15$ for  all the cases.

\section{Extension to General Environment} \label{sec:sumo}
In this section, we briefly discuss about the extension of the proposed scheme to the cases of multiple RB allocation at each user and time-varying available RBs in Section \ref{subsec:discussion}.
Throughout the paper, we have established RL-based dynamic multichannel access policies under the simplified setting in which throughput of one is achievable if there is no collision for a transmitted packet at each of RBs.
In practice, however, wireless channels can be affected by various physical phenomena such as path-loss, shadowing, and fading, which might also vary over time due to the mobility of users and/or scatters around them. 
As a result, an achievable throughput for successful packet transmission is determined by the corresponding channel gain.
In Sections \ref{subsec:wireless_channel} to \ref{subsec:numerical_SUMO}, we demonstrate that the proposed scheme is still applicable for the Shannon capacity based throughput metric and both throughput and fairness improvement is universally achievable for general practical time-varying channel environments.

\subsection{Discussion} \label{subsec:discussion}
\subsubsection{Multiple RB allocation at each user}
\begin{figure}[t] \centering
\includegraphics[scale=0.23]{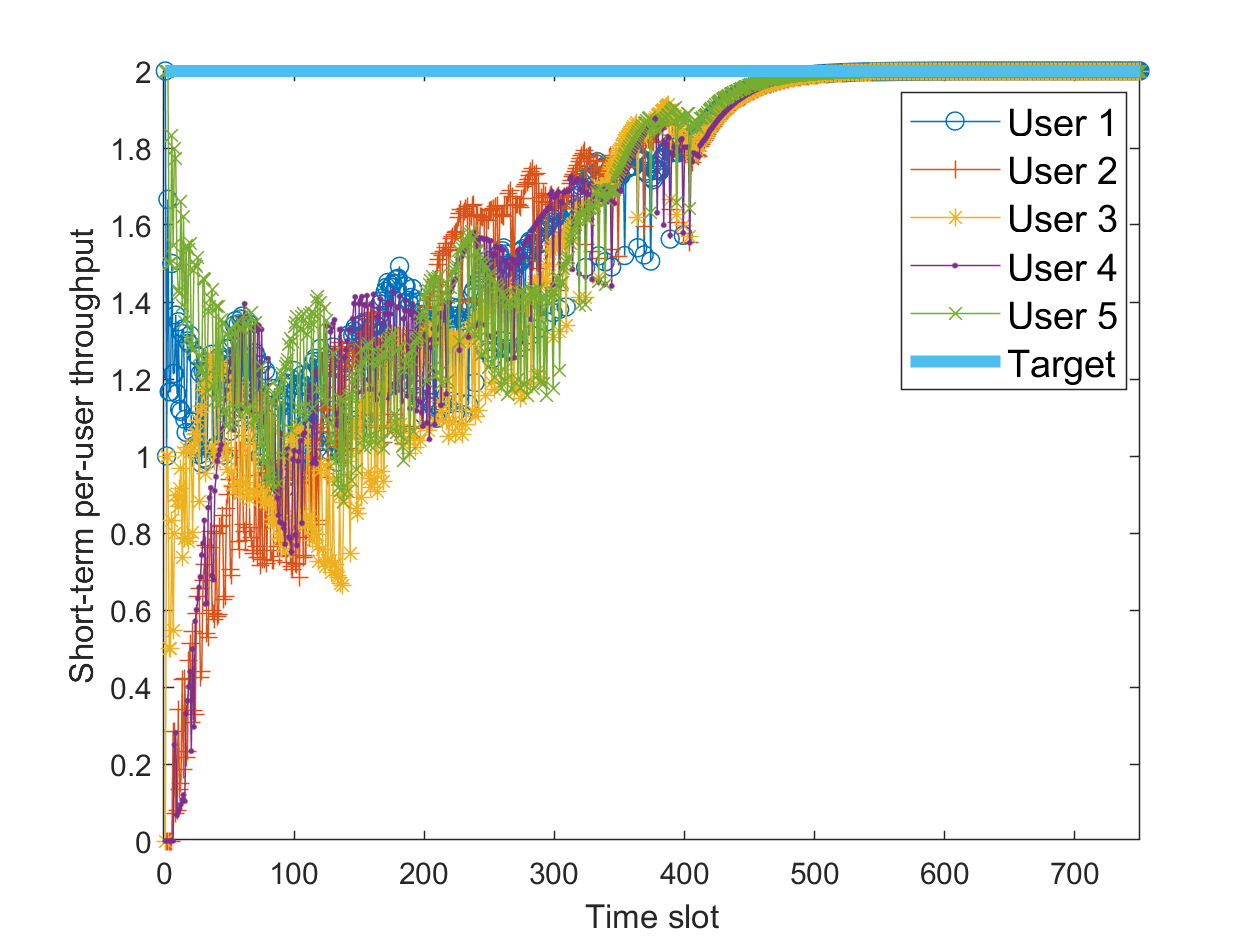}
\caption{$\Gamma_k(t)$  with respect to time slot $t$ for $k\in[1:K]$.}\label{fig:multple_RB_allocation}
\end{figure}

In this paper, RBs can be constructed in either frequency domain or time domain and single RB access is assumed at each user.
However, practical multichannel systems usually utilize multiple RBs in frequency domain like orthogonal frequency division multiplex (OFDM) and allow to allocate multiple RBs to a single user. 
It is worthwhile to mention that the proposed scheme can straightforwardly be extended to the case of allocating multiple RBs at each user. That is, each user selects $N$ RBs over $T[i]$ consecutive time slots and the transmit power should be properly divided for the time slots utilizing more than one RB. Fig. \ref{fig:multple_RB_allocation} plots $\Gamma_k(t)$ of the proposed scheme for $K=5$ and $N=10$, i.e., $\mathcal{K}(t)=[1:K]$ for all $t$, where we set $K_{\max}=K$ and $T_w=20$ in the figure. For this case, each user selects two RBs on average at each time slot and, as a result, each user's average throughput converges to two as the number of time slots increases. 

\subsubsection{Time-varying RBs}
\begin{figure}[H]
\begin{subfigure}{0.5\textwidth}
			\centering
         \includegraphics[width=3in]{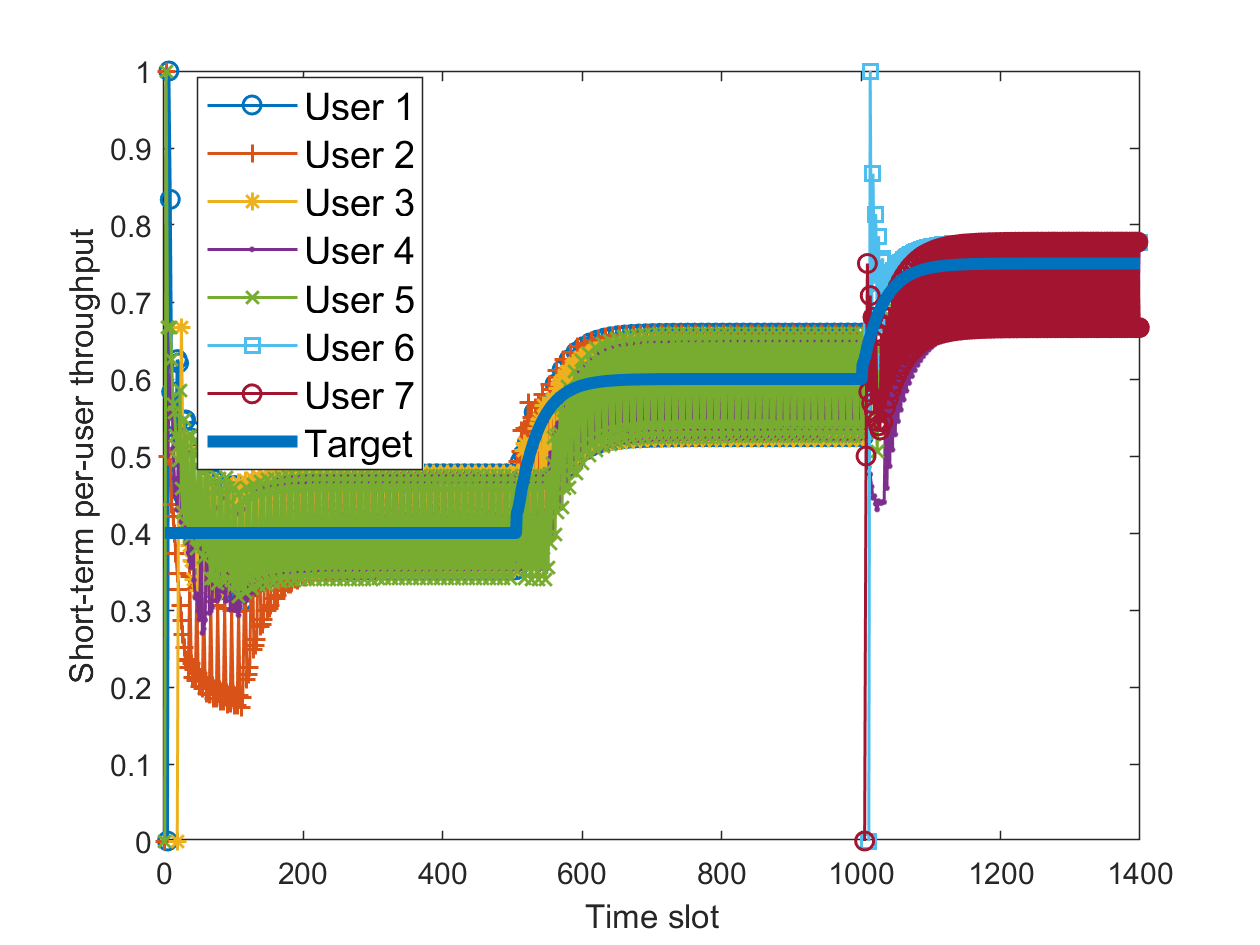}
         \caption{Short-term per-user throughputs}
     \end{subfigure}
     \begin{subfigure}{0.5\textwidth}
		\centering
         \includegraphics[width=3in]{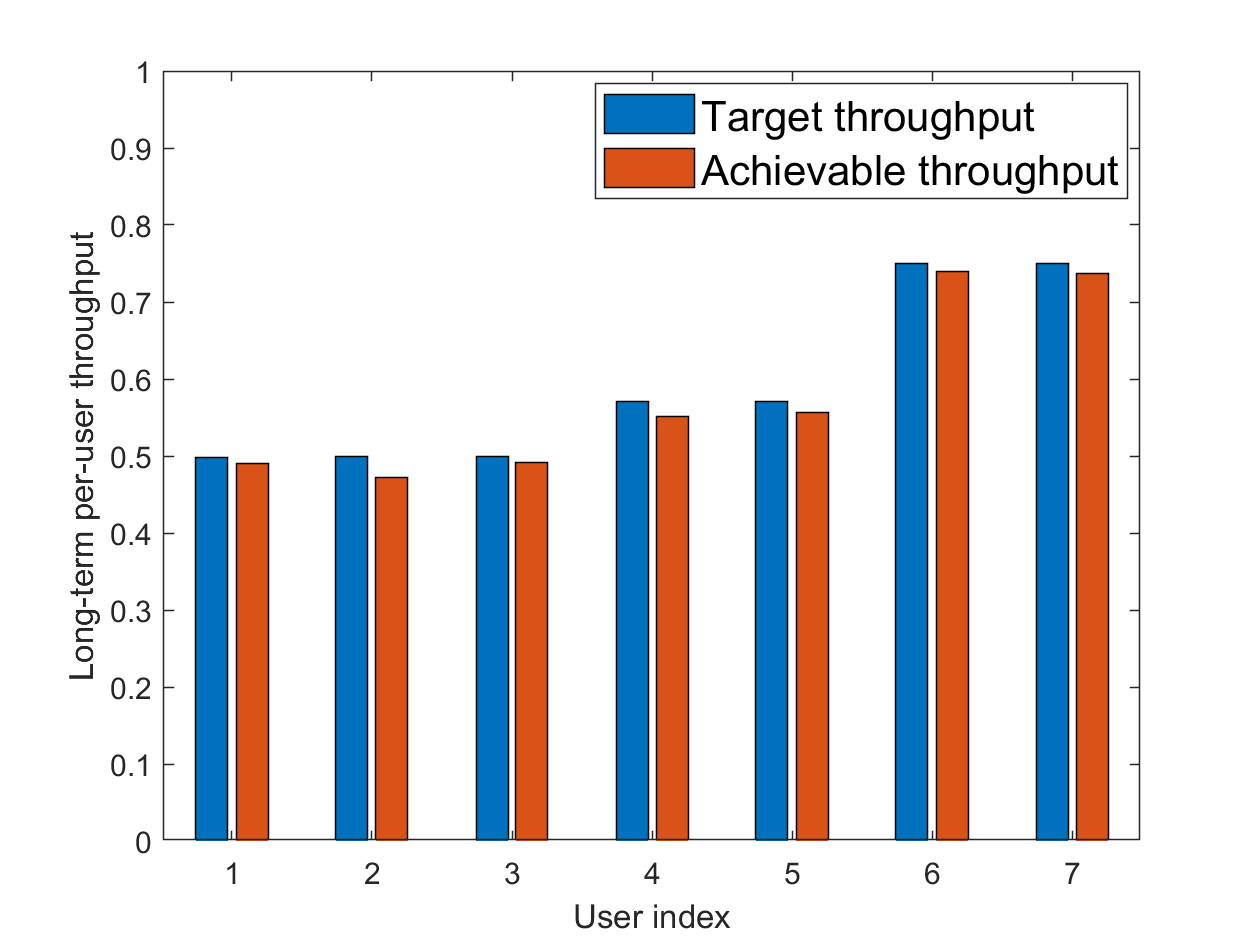}
         \caption{Long-term per-user throughputs}
         \label{fig:five over x}
     \end{subfigure}
        \caption{Short-term and long-term per-user throughputs for time-varying RBs and users.}
        \label{fig:varingchannels}
\end{figure}

For various reasons, the number of available RBs can be changed over time in practice. Then, in a similar way to the time-varying user case, we can introduce a predetermined value $N_{\max}$ and apply a vectorized Q-network and its training methodology.
Fig. \ref{fig:varingchannels} plots an example episode for short-term per-user throughputs of each active user with respect to time slots. At time slot $t=1$ , there are initially $5$ users and $2$ RBs and at time slot $t=500$ the number of users and RBs are changed to $5$ and $3$ respectively, and finally they are changed to $4$ and $3$ respectively at time slot $t=1000$. As seen in the figures, the proposed scheme can be easily generalized to scenarios with different available RBs.

\subsection{Wireless Channel Model and Throughput} \label{subsec:wireless_channel}
In this subsection, we introduce a general temporally and geometrically correlated wireless channels and the Shannon capacity based throughput metric.
In particular, we assume a circular-shaped cellular network with radius $\Delta$. The 2D position of AP is given by $\mathbf{u}_0=(0,0)$, which is fixed during the entire time slots.
We assume mobile users in which a user will enter the cell and traverse for a certain set of time slots and then go out the cell. Note that $t_{\operatorname{arr}, k}$ and  $t_{\operatorname{dep},k}$ are determined by the trajectory of user $k$. 
For $t\in [t_{\operatorname{arr}, k},t_{\operatorname{dep},k}]$, denote the 2D position of user $k$ at time slot $t$ by $\mathbf{u}_k(t)$, where $\|\mathbf{u}_k(t)\|\leq \Delta$.
The channel gain of RB $n$ and user $k$ at time slot $t\in [t_{\operatorname{arr}, k},t_{\operatorname{dep},k}]$ is given by
\begin{equation}
g_{k}^{n}(t)=\frac{1}{\|\mathbf{u}_k(t)\|^{\varrho}}|h_{k}^{n}(t)|^{2},
\end{equation}
where $\varrho\geq 2$ is the path-loss exponent and $h_{k}^{n}(t)$ denotes the small-scale channel coefficient of RB $n$ and user $k$ at time slot $t$. We adopt the Jake's model to represent the relationship between small-scale Rayleigh fading in two successive time slots by
\begin{equation}
h_{k}^{n}(t)=\xi h_{k}^{n}(t-1)+\delta,
\end{equation}
where $\xi\in[0,1)$ is the correlation coefficient  of two successive small-scale channel coefficients and $\delta$ is a random variable with a distribution $\mathcal{CN}(0,1-\xi^2)$. We also assume $h_{k}^{n}(0)$ follows $\mathcal{CN}(0,1)$.

An achievable throughput of user $k$ when it successfully sends a packet via RB $n$ at time slot $t$ is given by 
\begin{equation} \label{eq:shannon}
c_{k}^{n}(t)=\frac{W}{N}\log\left(1+\frac{g_{k}^{n}(t)P}{\frac{W}{N}N_{0}}\right),
\end{equation}
where $W$ is the system bandwidth, $P$ is the transmit power, and $N_0$ is the noise power spectral density. That is, the assigned bandwidth to each RB is given by $\frac{W}{N}$.

\subsection{Multi-agent RL Policy}
The proposed multi-agent RL policy in Sections \ref{sec:rl} and \ref{sec:DQN} is still applicable to general correlated wireless channels. In this subsection, we briefly explain the overall procedure of how to apply the proposed scheme for such environment.

Instead of the one--off throughput metric in \eqref{eq:ins_th}, an achievable throughput can be defined as
\begin{align} \label{eq:ins_th_new}
\gamma_k(t)=\begin{cases}c_{k}^{\eta_k(t)}(t)\mathbf{1}(b^{[\eta_k(t)]}(t)=\operatorname{ACK}) &\mbox{ if }\eta_k(t)\in[1:N],\\
0 &\mbox{ otherwise}
\end{cases}
\end{align}
from the Shannon capacity definition in \eqref{eq:shannon}. Then, from \eqref{eq:ins_th_new}, we define $\Gamma_k(t)$ as in \eqref{eq:th_ave}.
Furthermore, to measure fairness between the active users in $\mathcal{K}(t)$ in real-time, an upper bound on the instantaneous target per-user throughput can be defined as
\begin{align} \label{eq:th_target_new}
\gamma_{\operatorname{target},k}(t)=\min\left(1,\frac{N}{|\mathcal{K}(t)|}\right) \max_{n\in[1:N]} c_{k}^{n}(t). 
\end{align} 
From the above upper bound, $\Gamma_{\operatorname{target},k}(t)$ is defined as in \eqref{eq:th_target_ave}.
Finally, the short-term average throughput loss is defined as
\begin{align} \label{eq:Delta_k_modified}
\Delta_k=\frac{1}{T_k}\sum_{t=t_{\operatorname{arr}, k}}^{t_{\operatorname{dep},k}}\max\left(\frac{\Gamma_{\operatorname{target},k}(t)-\Gamma_k(t)}{\Gamma_{\operatorname{target},k}(t)},0\right).
\end{align}
Unlike the original definition in \eqref{eq:Delta_k}, we modified $\Delta_k$ as the loss ratio with respect to its target value so that the range of $\Delta_k$ in \eqref{eq:Delta_k_modified} is still between zero and one.

Recall that the main objective of each user at decision time $i$ , i.e., time slots $[T[i] : T[i] + K[i] -1]$, is to select a channel with the highest rate. Hence we re-define the reward of user $k$ at decision time $i$ as
 
\begin{align} \label{eq:reward_pf}
r_k[i]=\sum_{t=T[i]}^{T[i]+K[i]-1}r_k(t),
\end{align}
where
\begin{align} 
r_k(t)=\begin{cases}
1+\frac{c_{k}^{\eta_k(t)}(t)}{\max_{n\in[1:N]}c^n_{k}(t)} &\mbox{ if } b^{[\eta_k(t)]}(t)=\operatorname{ACK},\\
0 &\mbox{ if } \eta_k(t)=0,\\
-1+\frac{c_{k}^{\eta_k(t)}(t)}{\max_{n\in[1:N]}c^n_{k}(t)} &\mbox{ otherwise.}\\
\end{cases}
\end{align}
Similar to \eqref{eq:3step_reward}, we apply the three-step reward and add the rate based reward $\frac{c_{k}^{\eta_k(t)}(t)}{\max_{n\in[1:N]}c^n_{k}(t)}\in(0,1]$, which becomes one if the maximum rate channel is selected.

Finally, the state of user $k$ at decision time $i$ can be modified as
\begin{equation} \label{eq:state_new}
 \mathbf{s}_k[i]=(\mathbf{a}_{k}[i-1],\mathbf{r}_k[i-1],\{\mathbf{c}^n_k[i-1]\}_{n\in[1:N]}),
\end{equation}
where $\mathbf{c}_k^n[i]=\big[c^n_k(T[i]),\cdots, c^n_k(T[i]+K[i]-1)\big]$.
Because the re-defined throughput metrics are dependent on channel gains, $\{\mathbf{c}^n_k[i-1]\}_{n\in[1:N]}$ is obviously useful for each RL agent to maximize per-user throughputs.

For DQN training, the same methodology proposed in Section \ref{sec:DQN} is applicable. The only difference is that the size of state is changed in \eqref{eq:state_new} by adding $\{\mathbf{c}^n_k[i-1]\}_{n\in[1:N]}$ as its state information.
Hence, the input layer $\mathbf{x}$ in Fig. \ref{fig:DNN_architecture} should be changed accordingly.
In particular, let $\mathbf{x}\in\{0,1\}^{(N+1)K_{\max}}\times [-1,2]^{K_{\max}} \times  \mathbb{R}_+ ^{N}$, where $ \mathbb{R}_+$ denotes the set of positive real numbers. Then as the same manner in Section \ref{subsubsec:input},
For $K[i]=K_{\max}$, the input function is given by $\mathbf{x}=(\mathbf{a}'_k[i-1], \mathbf{r}_k[i-1],,\{\mathbf{c}^n_k[i-1]\}_{n\in[1:N]})$ at decision time $i$.
For $K[i]<K_{\max}$, the input function is given by $\mathbf{x}=(\mathbf{a}'_k[i-1], \mathbf{0}_{(N+1)(K_{\max}-|\mathcal{K}(T[i-1])|)}, \mathbf{r}_k[i-1], \mathbf{0}_{(K_{\max}-|\mathcal{K}(T[i-1])|)},,\{\mathbf{c}^n_k[i-1]\}_{n\in[1:N]})$ at decision time $i$. 

\subsection{Numerical Evaluation} \label{subsec:numerical_SUMO}

\begin{figure}[H]
      \begin{subfigure}{0.32\textwidth}
			\centering
         \includegraphics[width=\textwidth]{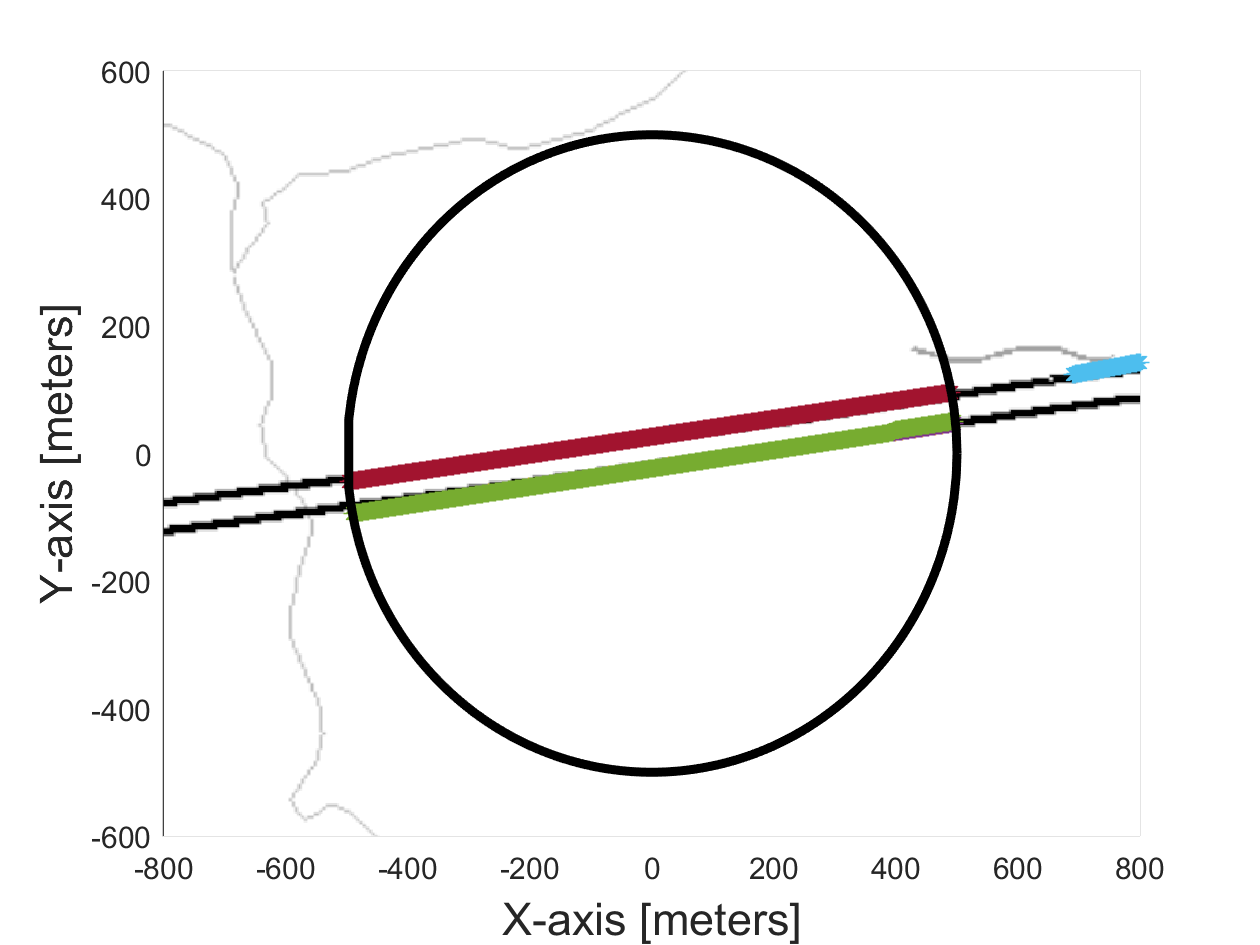}
         \caption{Highway}
     \end{subfigure}
     \begin{subfigure}{0.32\textwidth}
		\centering
         \includegraphics[width=\textwidth]{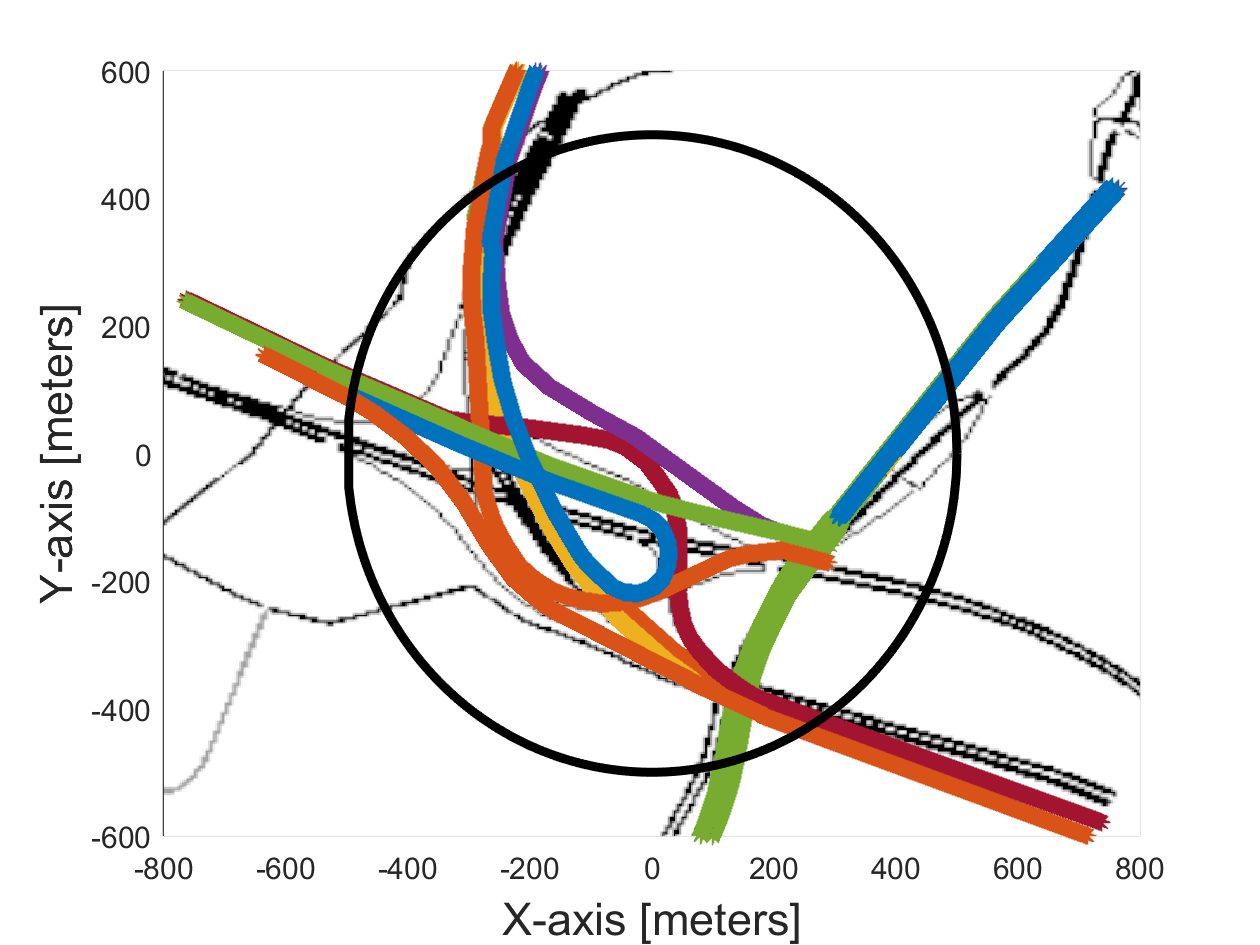}
         \caption{Highway intersection}
     \end{subfigure}
\begin{subfigure}{0.32\textwidth}
		\centering
         \includegraphics[width=\textwidth]{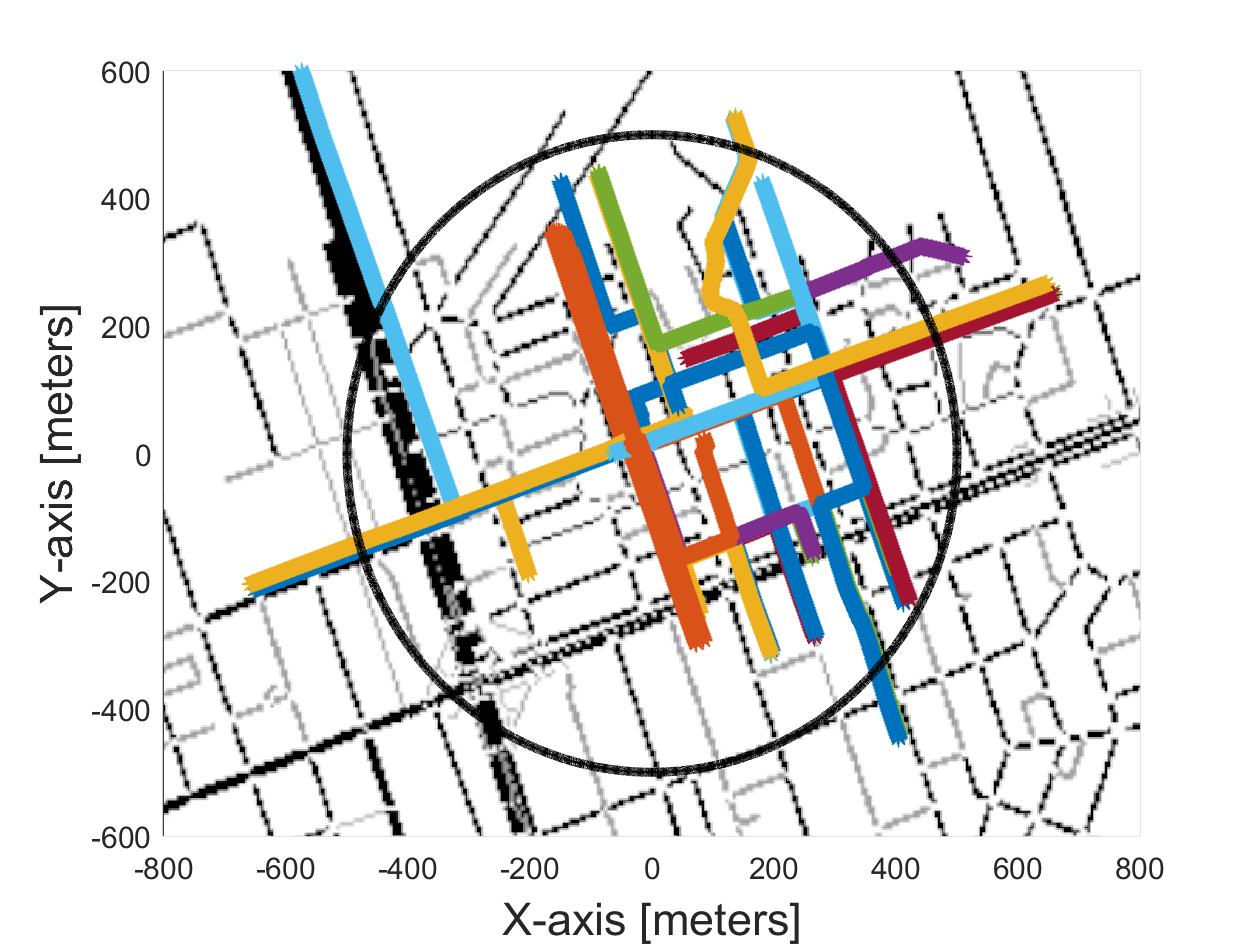}
         \caption{Urban}
     \end{subfigure} 
     \caption{Road maps used in three scenarios: (a) Ansan--Gunpo section, Seoul outer circulation expressway (37$^\circ$22'23.8" N 126$^\circ$54'51.5" E), (b)  Ansan junction (37$^\circ$20'39.2"N 126$^\circ$51'37.3" E), (c) Gangnam metro station, Seoul (37$^\circ$29'53.1" N 127$^\circ$01'40.5" E).} 
     \label{fig:maps}
\end{figure}

To reflect practical vehicular channel environments in numerical simulation, we apply the Simulation of Urban MObility (SUMO) simulator~\cite{51} using real road maps to generate temporal vehicular trajectories.
For performance evaluation, we consider three different scenarios: highway, highway intersection, and urban.
To represent the three cases, we choose real road maps in Korea illustrated in
Fig. \ref{fig:maps}. We also plot the cell radius and exemplary vehicular trajectories in the figures. Table \ref{tab:table10} summarizes the main simulation parameters.

\begin{table}[H]
  \begin{center}
    \caption{Simulation environment.}
    \label{tab:table10}
	\scalebox{0.85}{
    \begin{tabular}{l|c} 
     \hline 
	  \multicolumn{1}{c|}{Simulation parameters} & Values\\
      \hline\hline
      \multicolumn{1}{c|}{Path-loss exponent ($\varrho$)} & 3.38\\
		\hline
      \multicolumn{1}{c|}{Correlation coefficient of small fading ($\xi$)} & 0.9\\
		\hline
		\multicolumn{1}{c|}{Bandwidth ($W$)} & 20 MHz\\
		\hline
		\multicolumn{1}{c|}{Noise power spectral density ($N_0$)} & -174 dBm/Hz\\
       \hline
      \multicolumn{1}{c|}{Transmit power ($P$)}  & 23 dBm\\
		\hline
    \end{tabular}}
  \end{center}
\end{table}	 

For benchmark schemes, we consider two centralized schedulers selecting $\min(|\mathcal{K}(t)|,N)$ users in $\mathcal{K}(t)$ at each time slot $t$ by a central controller. The first benchmark scheme in Algorithm \ref{Algorithm4} chooses such users to maximize the sum throughput at each time slot in a greedy manner. 
For the second benchmark scheme, a PF metric is used for greedy user selection.
Let $r_{k, \operatorname{ave}}(t)=\frac{\sum_{i=\max(t_{\operatorname{arr},k},t-T_w)}^{t} \mathbf{c}_{k}^{\eta_{k}(i)}(i)}{t-\max(t_{\operatorname{arr},k},t-T_w)+1}$ be the average throughput of user $k$ up to time slot $t$. Then define $c_{k, \operatorname{PF}}^n(t)=\frac{c^n_k(t)}{r_{k, \operatorname{ave}}(t-1)}$, which is a PF metric for greedy user selection.
Hence, the same scheme in  Algorithm \ref{Algorithm4} can be used for the second benchmark scheme by replacing $\{c_{k}^{n}(t) \}_{k \in \mathcal{K}(t), n \in [1:N]} $ with $\{c_{k, \operatorname{PF}}^n(t) \}_{k \in \mathcal{K}(t), n \in [1:N]} $.

\begin{algorithm}[] \footnotesize
\caption{Centralized scheduling.}\label{Algorithm4}
\begin{algorithmic}[3]
\State{{\bf Input}: $\mathcal{K}(t)$ and $\{c_{k}^{n}(t) \}_{k \in \mathcal{K}(t), n \in [1:N]} $.}
\State{{\bf Initialization}: Define $\mathcal{K}=\mathcal{K}(t)$ and $\mathcal{N}=[1:N]$. Set $\eta_k(t)=0$ for all $k\in\mathcal{K}(t)$.}
\For{$i=1$ to $\min(|\mathcal{K}(t)|,N)$}
\State {Calculate $(k^*, n^*)=\mathop {\arg\max }\limits_{k\in\mathcal{K},n\in\mathcal{N}}[c_{k}^{n}(t)]$.}
\State{Update  $\eta_{k^{*}}(t)\leftarrow n^*$, $\mathcal{K}\leftarrow \mathcal{K}\setminus\{k^*\}$, and $\mathcal{N}\leftarrow \mathcal{N}\setminus\{n^*\}$.}
\EndFor
\State{{\bf Output}: $\{ \eta_{k}(t) \}_{k \in \mathcal{K}(t) }$.}
 \end{algorithmic}
\end{algorithm}

\begin{figure}[]
     \begin{subfigure}{0.5\textwidth}
			\centering
         \includegraphics[width=\textwidth]{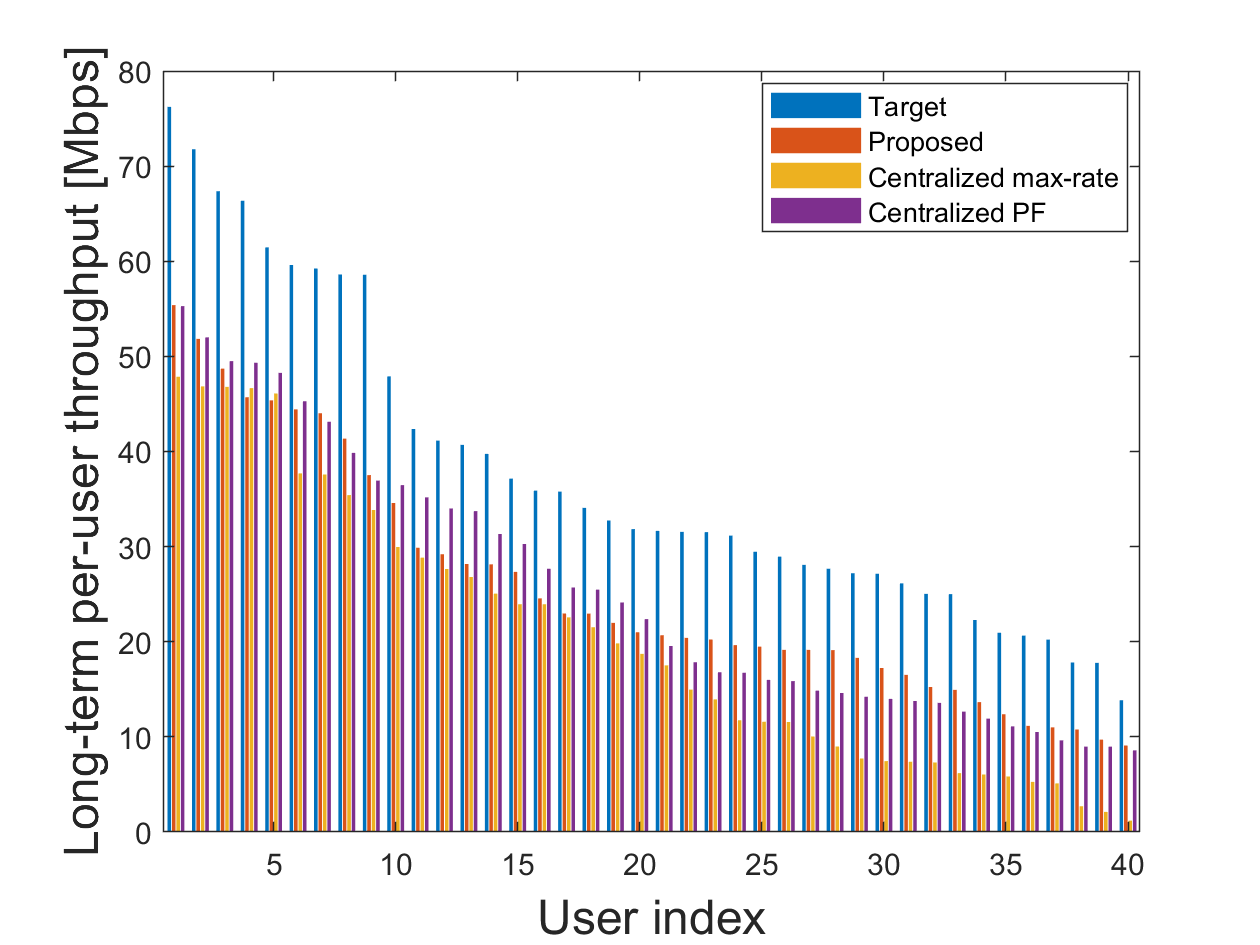}
         \caption{Highway, $\lambda=0.035$, $N=5$ }
     \end{subfigure}
     \begin{subfigure}{0.5\textwidth}
		\centering
         \includegraphics[width=\textwidth]{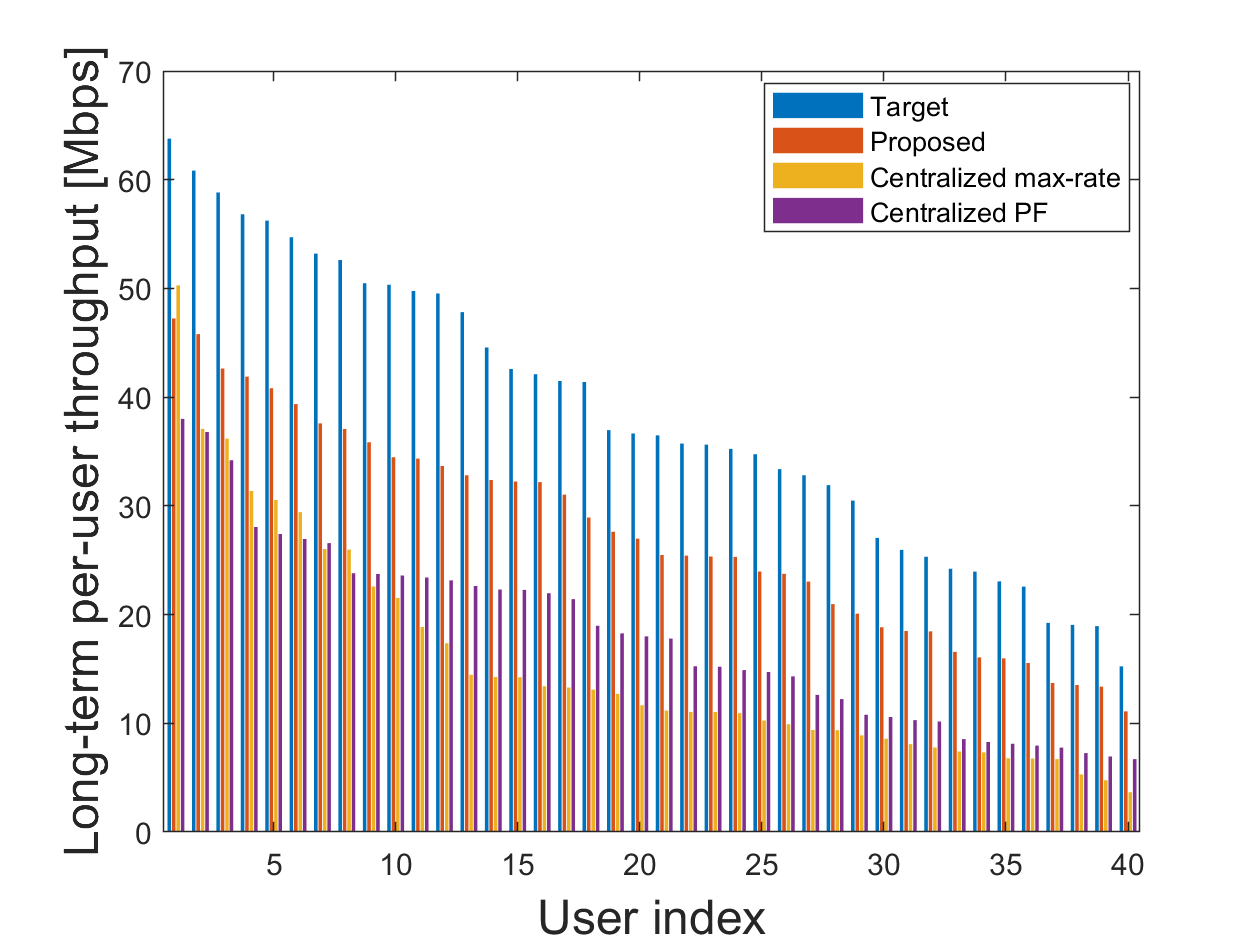}
		  \caption{Highway intersection, $\lambda=0.035$, $N=5$  }
     \end{subfigure}
     \begin{subfigure}{0.5\textwidth}
		 \centering
         \includegraphics[width=\textwidth]{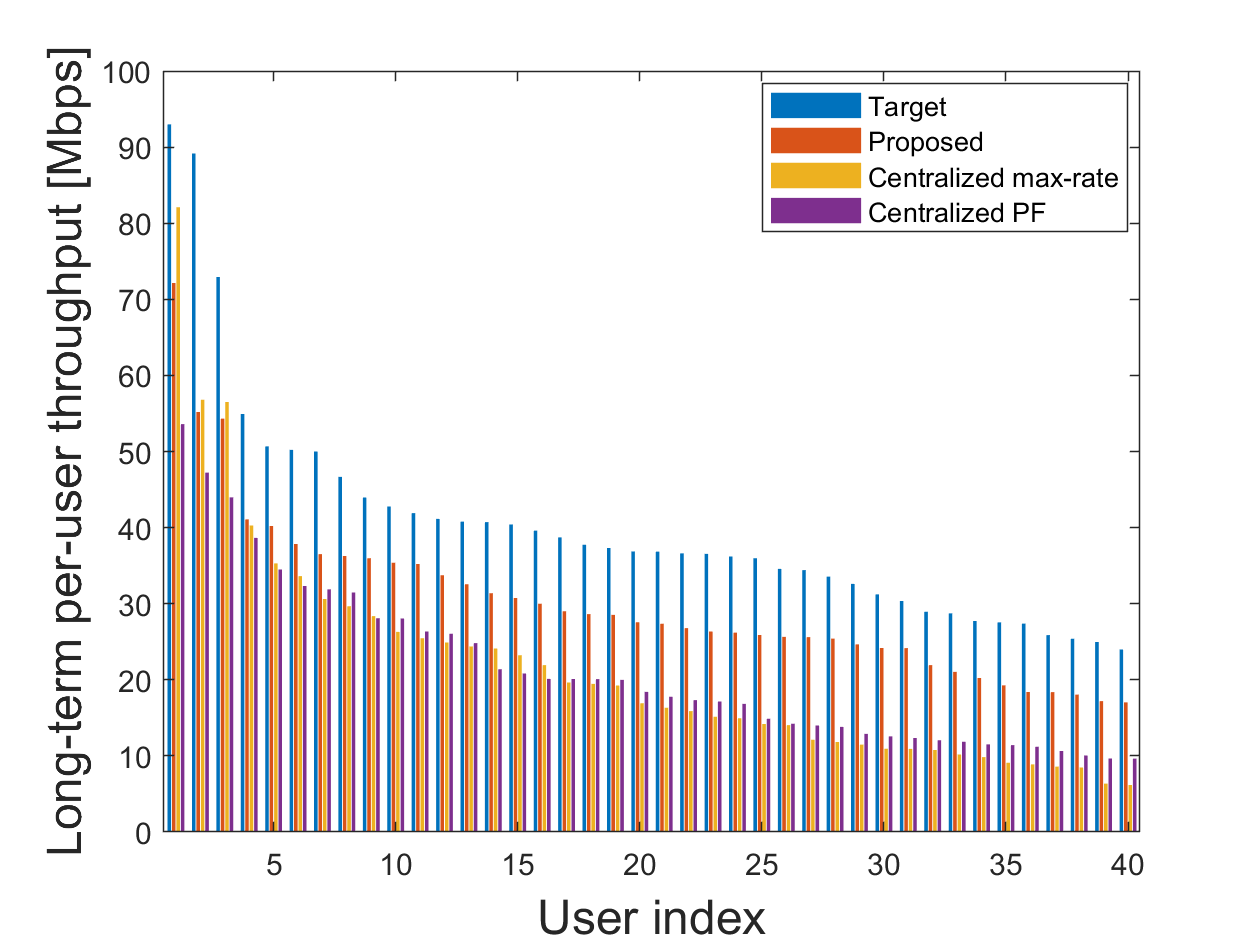}
         \caption{Urban, $\lambda=0.035$, $N=5$  }
     \end{subfigure}
     \begin{subfigure}{0.5\textwidth}
		 \centering
         \includegraphics[width=\textwidth]{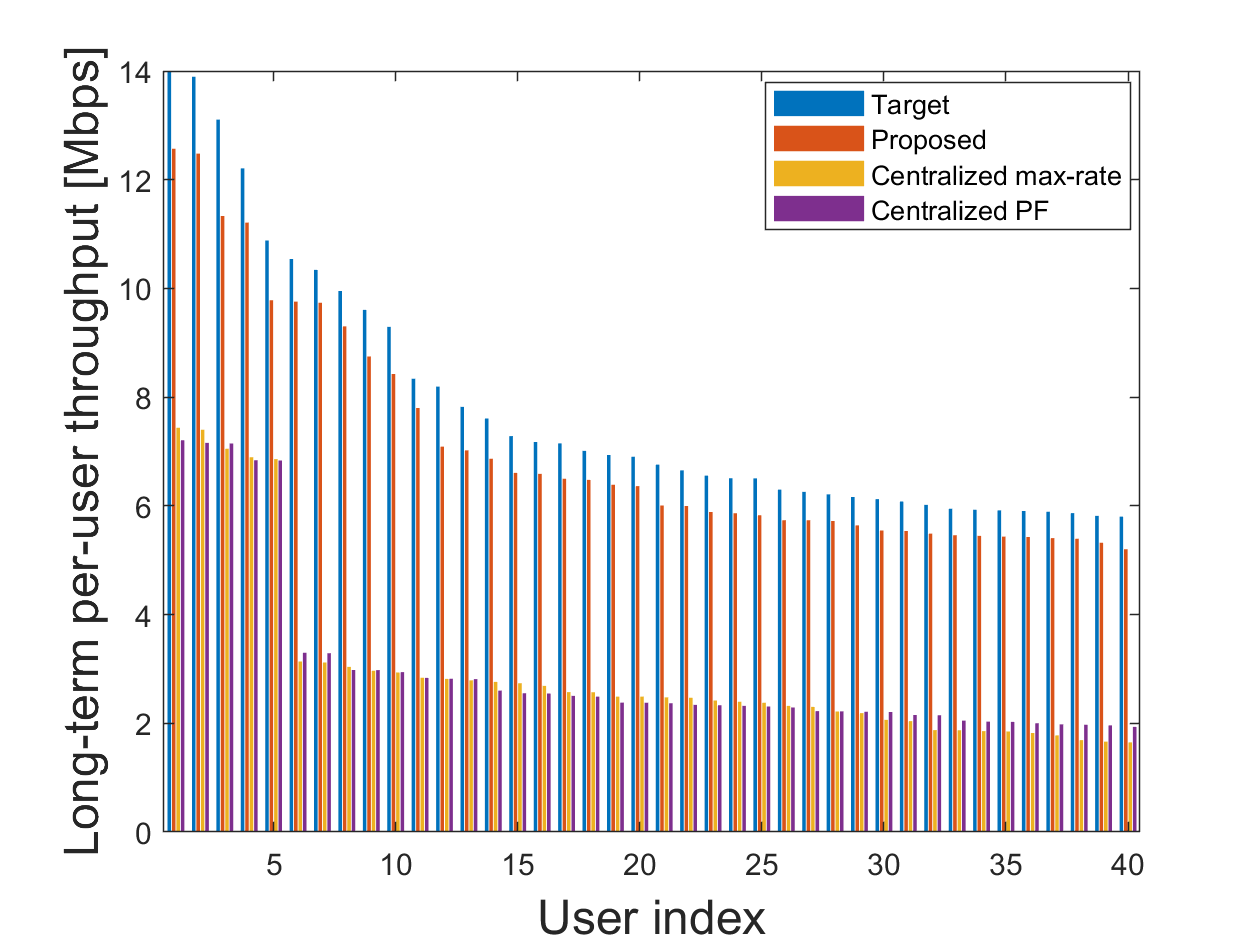}
         \caption{Urban, $\lambda=0.2$, $N=20$  }
     \end{subfigure}
\caption{Long-term per-user throughputs.} \label{fig:long_term_sumo}
\end{figure}
\begin{figure}[]
     \begin{subfigure}{0.5\textwidth}
			\centering
         \includegraphics[width=0.9\textwidth]{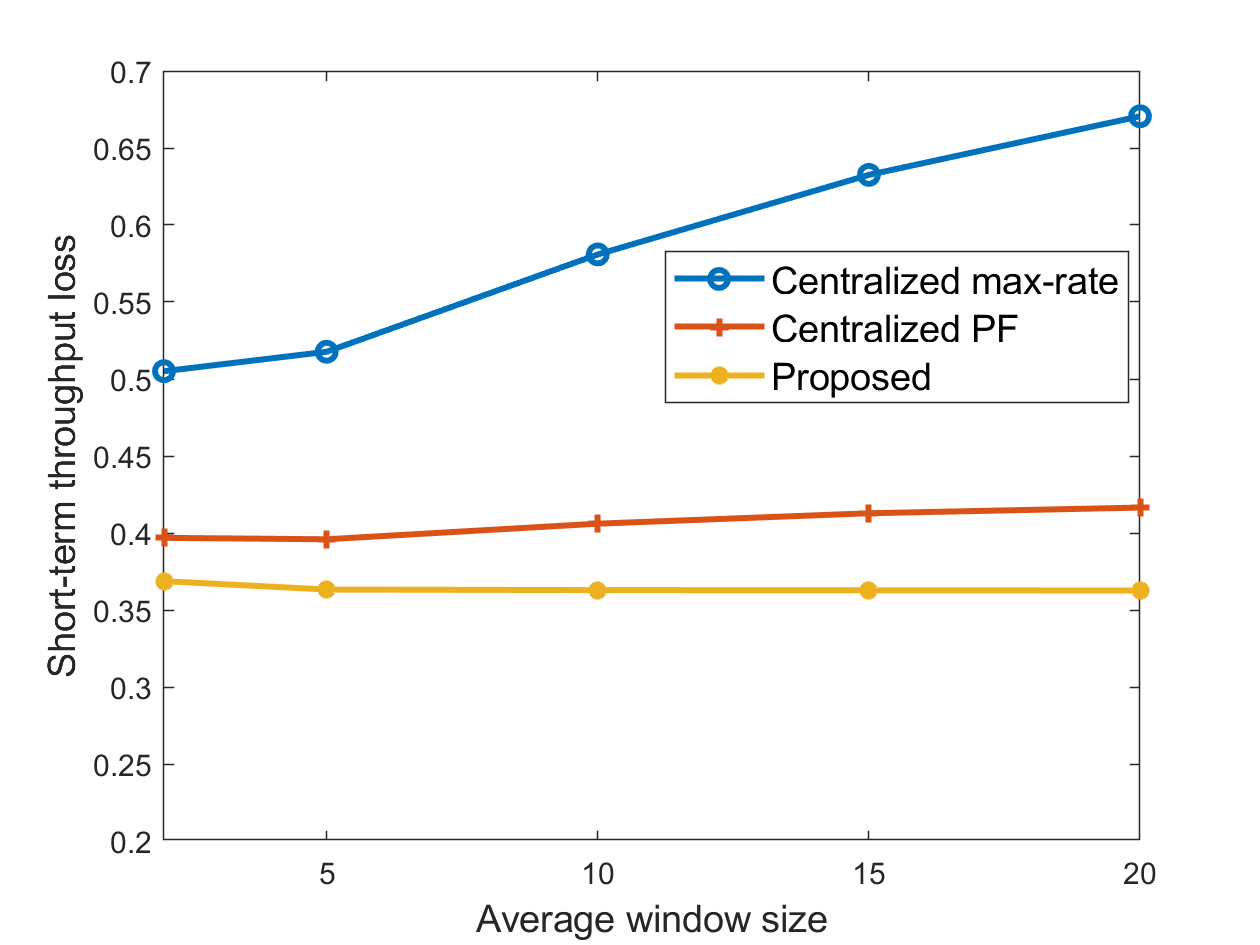}
         \caption{Highway, $\lambda=0.035$, $N=5$ }
     \end{subfigure}
     \begin{subfigure}{0.5\textwidth}
		\centering
         \includegraphics[width=0.9\textwidth]{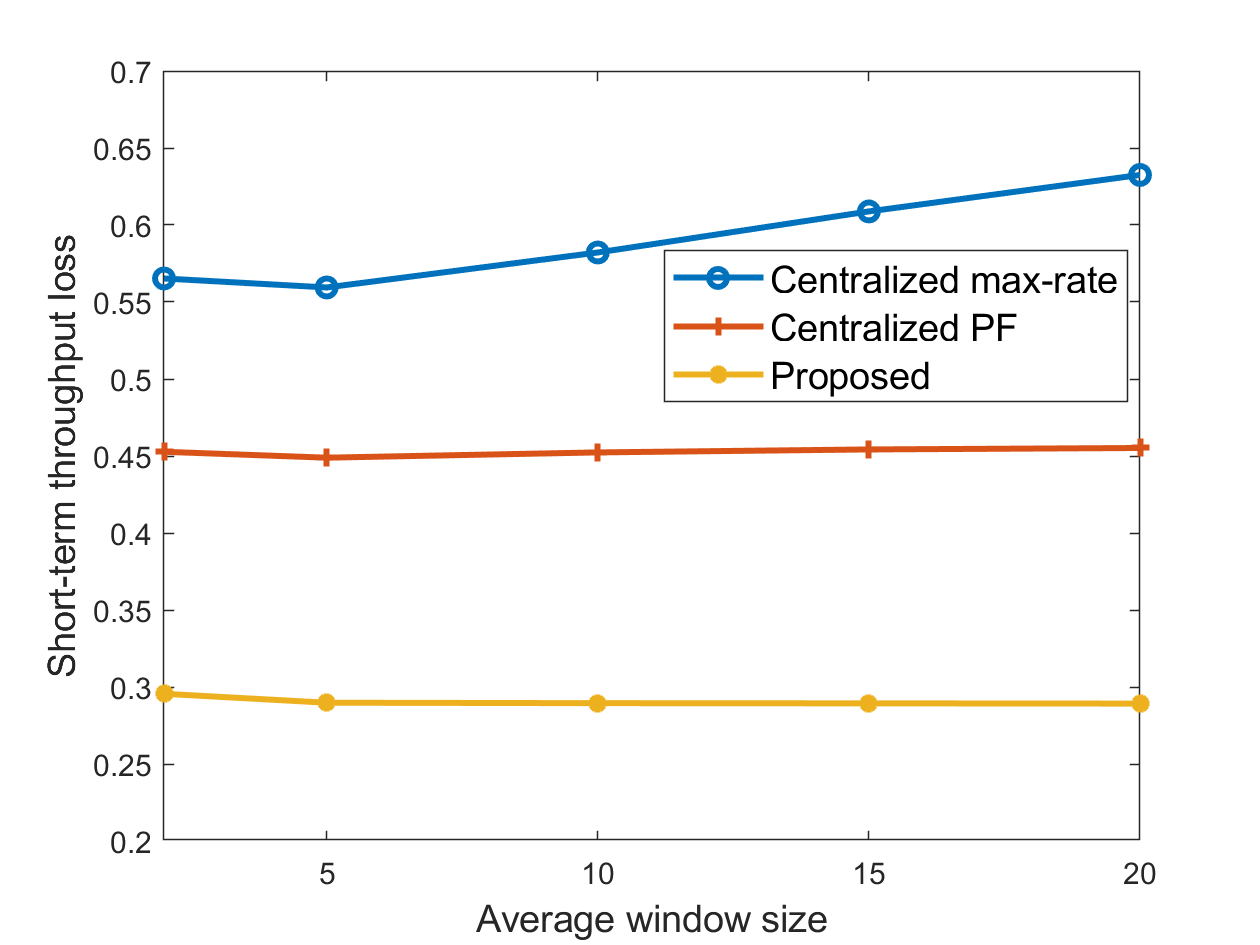}
		  \caption{Highway intersection, $\lambda=0.035$, $N=5$ }
     \end{subfigure}
     \begin{subfigure}{0.5\textwidth}
		 \centering
         \includegraphics[width=0.9\textwidth]{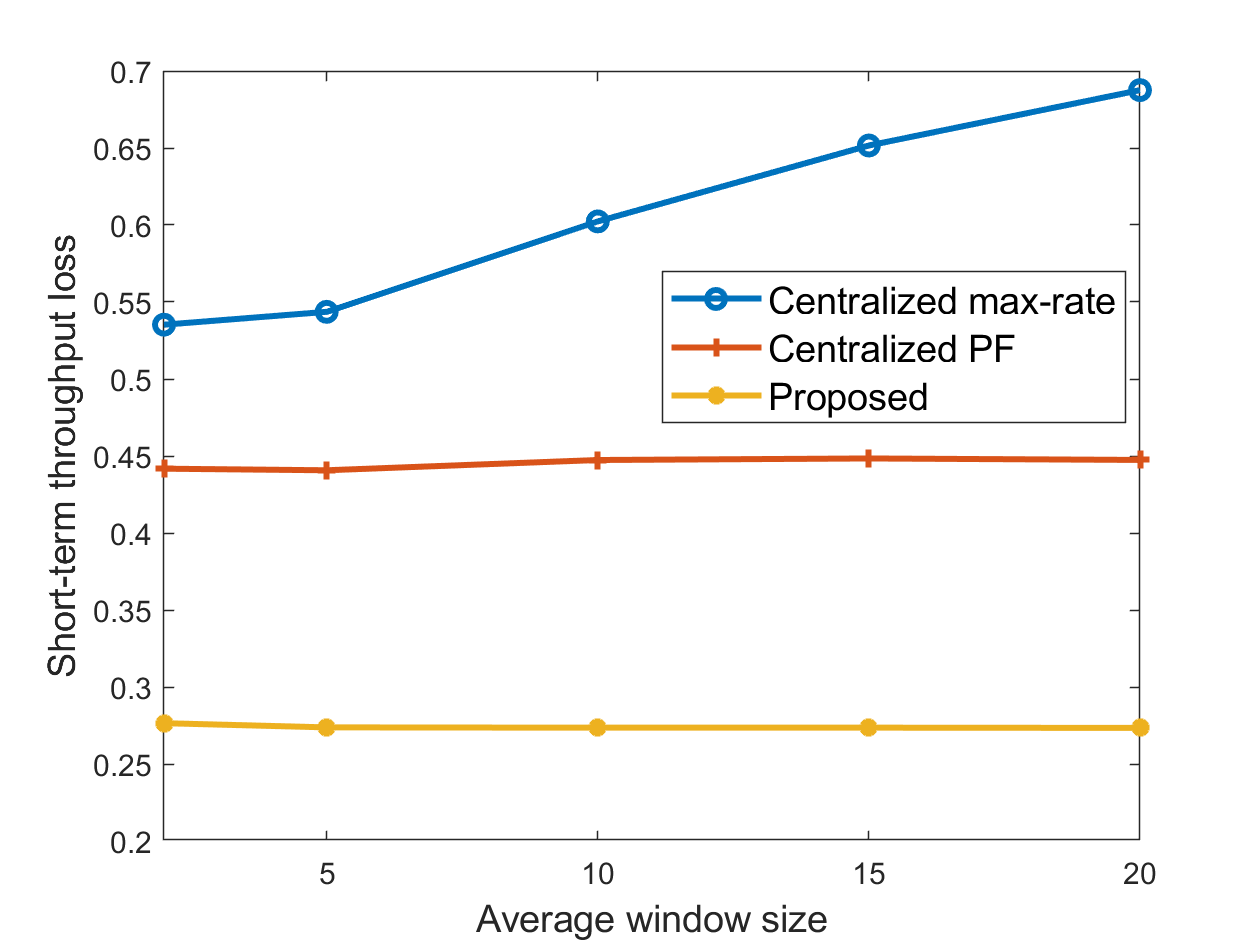}
         \caption{Urban, $\lambda=0.035$, $N=5$}
     \end{subfigure}
     \begin{subfigure}{0.5\textwidth}
		 \centering
         \includegraphics[width=0.9\textwidth]{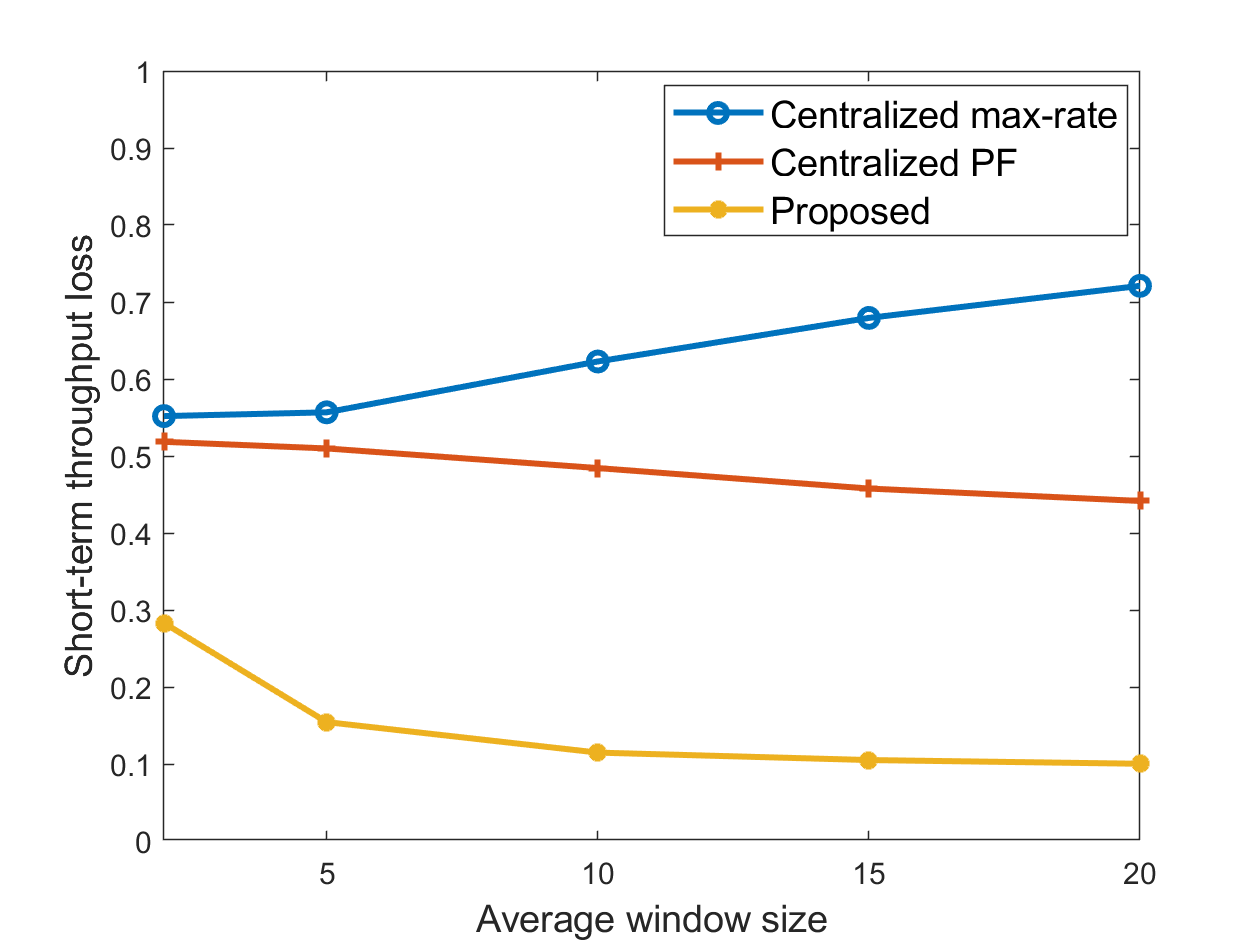}
         \caption{Urban, $\lambda=0.2$, $N=20$}
     \end{subfigure}
\caption{$\Delta_k$ with respect to the average window size $T_w$.} \label{fig:short_term_sumo}
\end{figure}

Fig. \ref{fig:long_term_sumo} plots long-term average per-user throughput of the proposed scheme along with two centralized user scheduling schemes for three different environments when $\lambda=0.035$ and $N=5$ (Fig. \ref{fig:long_term_sumo} (a), (b), and (c) ) and when $\lambda=0.2$ and $N=20$ (Fig. \ref{fig:long_term_sumo} (d) ). Table \ref{tab:table 5} states the long-term average sum throughputs for three different environments when $\lambda=0.035$ and $N=5$ and $\lambda=0.2$ and $N=20$.
Note that the proposed scheme even provides an improved per-user and sum throughputs compared to the centralized schemes. To check the short-term fairness performance, Fig. \ref{fig:short_term_sumo} plots $\Delta_k$ in \eqref{eq:Delta_k_modified} with respect to $T_w$. Not only improving long-term throughputs, but the proposed scheme also provides better short-term fairness between active users.

\begin{table}\caption{Long-term average sum throughput.}\label{tab:table 5}
\begin{center}
\scalebox{0.85}{
\begin{tabular}{c|c|c|c|c|}
\hline
\multirow[|c]{3}{*}{Schemes} & \multicolumn{4}{c}{Environments}\\
\cline{2-5}
&  Highway & Highway intersection & \multicolumn{1}{c|}{Urban} & \multicolumn{1}{c}{Urban}\\
\cline{2-5}
& $\lambda=0.035$, $N=5$ & $\lambda=0.035$, $N=5$ & $\lambda=0.035$, $N=5$ & \multicolumn{1}{c}{$\lambda=0.2$, $N=20$} \\
\cline{2-5}
\hline
\multirow[|c]{1}{*}{Proposed} & {$131.0$ Mbps} & {$96.6$ Mbps} & \multicolumn{1}{c|}{$123.0$ Mbps} & \multicolumn{1}{c}{$190.0$ Mbps} \\
\hline
\multirow[|c]{1}{*}{Centralized max-rate} & {$90.1$ Mbps} & {$42.0$ Mbps} & \multicolumn{1}{c|}{$53.9$ Mbps}& \multicolumn{1}{c}{$127.0$ Mbps}\\
\hline
\multirow[|c]{1}{*}{Centralized PF} & {$69.6$ Mbps} & {$40.9$ Mbps} & \multicolumn{1}{c|}{$65.1$ Mbps} & \multicolumn{1}{c}{$128.4$ Mbps}\\
\hline
\end{tabular}}
\end{center}
\end{table}

\section{Conclusion} \label{sec:conclusion}
In this paper, we considered a dynamic multichannel access problem to guarantee both throughput and fairness under dynamic network environment. Numerical results showed that the proposed RL algorithm provides an improved real-time fairness between active user while preserving throughput for each user.
Our work can be extended to several interesting further research directions, for example, joint optimization considering user association for multi-cell environment or transmit power adaptation would be a promising future work. Although we assumed an orthogonal multichannel network sharing $N$ orthogonal RBs between active users, the proposed deep neural network structure and training methodologies are applicable to NOMA systems with dynamic user environment.




\begin{thebibliography}{10}
\providecommand{\url}[1]{#1}
\csname url@samestyle\endcsname
\providecommand{\newblock}{\relax}
\providecommand{\bibinfo}[2]{#2}
\providecommand{\BIBentrySTDinterwordspacing}{\spaceskip=0pt\relax}
\providecommand{\BIBentryALTinterwordstretchfactor}{4}
\providecommand{\BIBentryALTinterwordspacing}{\spaceskip=\fontdimen2\font plus
\BIBentryALTinterwordstretchfactor\fontdimen3\font minus
  \fontdimen4\font\relax}
\providecommand{\BIBforeignlanguage}[2]{{%
\expandafter\ifx\csname l@#1\endcsname\relax
\typeout{** WARNING: IEEEtran.bst: No hyphenation pattern has been}%
\typeout{** loaded for the language `#1'. Using the pattern for}%
\typeout{** the default language instead.}%
\else
\language=\csname l@#1\endcsname
\fi
#2}}
\providecommand{\BIBdecl}{\relax}
\BIBdecl

\bibitem{12}
N.~C. {Luong}, D.~T. {Hoang}, S.~{Gong}, D.~{Niyato}, P.~{Wang}, Y.~{Liang},
  and D.~I. {Kim}, ``Applications of deep reinforcement learning in
  communications and networking: A survey,'' \emph{{IEEE} Commun. Surveys
  Tuts.}, vol.~21, no.~4, pp. 3133--3174, 2019.

\bibitem{22}
Y.~{Zhou}, Z.~M. {Fadlullah}, B.~{Mao}, and N.~{Kato}, ``A deep-learning-based
  radio resource assignment technique for {5G} ultra dense networks,''
  \emph{IEEE Network}, vol.~32, no.~6, pp. 28--34, 2018.

\bibitem{41}
E.~{De Carvalho}, E.~{Bjornson}, J.~H. {Sorensen}, P.~{Popovski}, and E.~G.
  {Larsson}, ``Random access protocols for massive {MIMO},'' \emph{IEEE
  Communications Magazine}, vol.~55, no.~5, pp. 216--222, 2017.

\bibitem{42}
R.~{Harwahyu}, R.~{Cheng}, C.~{Wei}, and R.~F. {Sari}, ``Optimization of random
  access channel in {NB-IoT},'' \emph{IEEE Internet of Things Journal}, vol.~5,
  no.~1, pp. 391--402, 2018.

\bibitem{9260174}
B.~H. {Lee}, H.~S. {Lee}, S.~{Moon}, and J.~W. {Lee}, ``Enhanced random access
  for {M}assive-{M}achine-{T}ype communications,'' \emph{IEEE Internet of
  Things Journal}, vol.~8, no.~8, pp. 7046--7064, 2021.

\bibitem{43}
J.~{Capetanakis}, ``Tree algorithms for packet broadcast channels,''
  \emph{{IEEE} Trans. Inf. Theory}, vol.~25, no.~5, pp. 505--515, 1979.

\bibitem{45}
M.~{Sidi} and I.~{Cidon}, ``Splitting protocols in presence of capture,''
  \emph{{IEEE} Trans. Inf. Theory}, vol.~31, no.~2, pp. 295--301, 1985.

\bibitem{46}
G.~{Fayolle}, P.~{Flajolet}, M.~{Hofri}, and P.~{Jacquet}, ``Analysis of a
  stack algorithm for random multiple-access communication,'' \emph{{IEEE}
  Trans. Inf. Theory}, vol.~31, no.~2, pp. 244--254, 1985.

\bibitem{47}
Y.~{Yu} and G.~B. {Giannakis}, ``High-throughput random access using successive
  interference cancellation in a tree algorithm,'' \emph{{IEEE} Trans. Inf.
  Theory}, vol.~53, no.~12, pp. 4628--4639, 2007.

\bibitem{jeon}
S.-W. Jeon and H.~Jin, ``Online estimation and adaptation for random access
  with successive interference cancellation,'' in \emph{IEEE International
  Symposium on Information Theory {(ISIT)}}, 2020.

\bibitem{48}
S.~{Lee}, B.~C. {Jung}, and S.-W. {Jeon}, ``Successive interference
  cancellation with feedback for random access networks,'' \emph{{IEEE} Commun.
  Lett.}, vol.~21, no.~4, pp. 825--828, 2017.

\bibitem{CHU201523}
Y.~Chu, S.~Kosunalp, P.~D. Mitchell, D.~Grace, and T.~Clarke, ``Application of
  reinforcement learning to medium access control for wireless sensor
  networks,'' \emph{Engineering Applications of Artificial Intelligence},
  vol.~46, pp. 23--32, 2015.

\bibitem{17}
S.~{Wang}, H.~{Liu}, P.~H. {Gomes}, and B.~{Krishnamachari}, ``Deep
  reinforcement learning for dynamic multichannel access in wireless
  networks,'' \emph{IEEE Trans. Cogn. Commun. Netw.}, vol.~4, no.~2, pp.
  257--265, 2018.

\bibitem{28}
C.~{Zhong}, Z.~{Lu}, M.~C. {Gursoy}, and S.~{Velipasalar}, ``A deep
  actor-critic reinforcement learning framework for dynamic multichannel
  access,'' \emph{IEEE Trans. Cogn. Commun. Netw.}, vol.~5, no.~4, pp.
  1125--1139, 2019.

\bibitem{3}
T.~{Gafni} and K.~{Cohen}, ``A distributed stable strategy learning algorithm
  for multi-user dynamic spectrum access,'' in \emph{Annual Allerton Conference
  on Communication, Control, and Computing (Allerton)}, 2019, pp. 347--351.

\bibitem{26}
H.~{Yang}, X.~{Xie}, and M.~{Kadoch}, ``Intelligent resource management based
  on reinforcement learning for ultra-reliable and low-latency {IoV}
  communication networks,'' \emph{{IEEE} Trans. Veh. Technol.}, vol.~68, no.~5,
  pp. 4157--4169, 2019.

\bibitem{27}
L.~{Liang}, H.~{Ye}, and G.~Y. {Li}, ``Spectrum sharing in vehicular networks
  based on multi-agent reinforcement learning,'' \emph{{IEEE} J. Select. Areas
  Commun.}, vol.~37, no.~10, pp. 2282--2292, 2019.

\bibitem{6}
N.~{Zhao}, Y.~{Liang}, D.~{Niyato}, Y.~{Pei}, M.~{Wu}, and Y.~{Jiang}, ``Deep
  reinforcement learning for user association and resource allocation in
  heterogeneous cellular networks,'' \emph{{IEEE} Trans. Wireless Commun.},
  vol.~18, no.~11, pp. 5141--5152, 2019.

\bibitem{38}
Y.~{Yu}, T.~{Wang}, and S.~C. {Liew}, ``Deep-reinforcement learning multiple
  access for heterogeneous wireless networks,'' \emph{{IEEE} J. Select. Areas
  Commun.}, vol.~37, no.~6, pp. 1277--1290, 2019.

\bibitem{21}
V.~{Raj}, I.~{Dias}, T.~{Tholeti}, and S.~{Kalyani}, ``Spectrum access in
  cognitive radio using a two-stage reinforcement learning approach,''
  \emph{{IEEE} J. Sel. Topics Signal Process.}, vol.~12, no.~1, pp. 20--34,
  2018.

\bibitem{20}
X.~{Li}, J.~{Fang}, W.~{Cheng}, H.~{Duan}, Z.~{Chen}, and H.~{Li},
  ``Intelligent power control for spectrum sharing in cognitive radios: A deep
  reinforcement learning approach,'' \emph{IEEE Access}, vol.~6, pp.
  25\,463--25\,473, 2018.

\bibitem{34}
H.~{Chang}, H.~{Song}, Y.~{Yi}, J.~{Zhang}, H.~{He}, and L.~{Liu},
  ``Distributive dynamic spectrum access through deep reinforcement learning: A
  reservoir computing-based approach,'' \emph{IEEE Internet of Things Journal},
  vol.~6, no.~2, pp. 1938--1948, 2019.

\bibitem{9352956}
J.~{Zheng}, X.~{Tang}, X.~{Wei}, H.~{Shen}, and L.~{Zhao}, ``Channel assignment
  for hybrid {NOMA} systems with deep reinforcement learning,'' \emph{{IEEE}
  Commun. Lett.}, {Early Access}.

\bibitem{9102308}
M.~{Nduwayezu}, Q.~{Pham}, and W.~{Hwang}, ``Online computation offloading in
  {NOMA}-based multi-access edge computing: A deep reinforcement learning
  approach,'' \emph{IEEE Access}, vol.~8, pp. 99\,098--99\,109, 2020.

\bibitem{8952876}
N.~{Ye}, X.~{Li}, H.~{Yu}, L.~{Zhao}, W.~{Liu}, and X.~{Hou}, ``Deep{NOMA}: A
  unified framework for {NOMA} using deep multi-task learning,'' \emph{{IEEE}
  Trans. Wireless Commun.}, vol.~19, no.~4, pp. 2208--2225, 2020.

\bibitem{23}
S.~M. {Zafaruddin}, I.~{Bistritz}, A.~{Leshem}, and D.~{Niyato}, ``Distributed
  learning for channel allocation over a shared spectrum,'' \emph{{IEEE} J.
  Select. Areas Commun.}, vol.~37, no.~10, pp. 2337--2349, 2019.

\bibitem{8}
K.~{Kar}, S.~{Sarkar}, and L.~{Tassiulas}, ``Achieving proportional fairness
  using local information in aloha networks,'' \emph{IEEE Trans. on Automatic
  Control}, vol.~49, no.~10, pp. 1858--1863, 2004.

\bibitem{9}
{H. Zheng} and {C. Peng}, ``Collaboration and fairness in opportunistic
  spectrum access,'' in \emph{IEEE International Conference on Communications
  {(ICC)}}, 2005, pp. 3132--3136.

\bibitem{40}
Y.~S. {Nasir} and D.~{Guo}, ``Multi-agent deep reinforcement learning for
  dynamic power allocation in wireless networks,'' \emph{{IEEE} J. Select.
  Areas Commun.}, vol.~37, no.~10, pp. 2239--2250, 2019.

\bibitem{11}
O.~{Naparstek} and K.~{Cohen}, ``Deep multi-user reinforcement learning for
  distributed dynamic spectrum access,'' \emph{{IEEE} Trans. Wireless Commun.},
  vol.~18, no.~1, pp. 310--323, 2019.

\bibitem{25}
I.~{Budhiraja}, N.~{Kumar}, and S.~{Tyagi}, ``Deep reinforcement learning based
  proportional fair scheduling control scheme for underlay {D2D}
  communication,'' \emph{IEEE Internet of Things Journal}, {Early Access}.

\bibitem{44}
D.~P. Bertsekas and R.~Gallager, \emph{Data networks, Pentice-Hall, 1987}.

\bibitem{49}
R.~S. Sutton and A.~G. Barto, \emph{Reinforcement learning: An
  introduction}.\hskip 1em plus 0.5em minus 0.4em\relax MIT press, 2018.

\bibitem{52}
V.~{Mnih}, K.~{Kavukcuoglu}, D.~{Silver}, A.~{Graves}, I.~{Antonoglou},
  D.~{Wierstra}, and M.~{Riedmiller}, ``Playing atari with deep reinforcement
  learning,'' \emph{arXiv preprint arXiv:1312.5602}, 2013.

\bibitem{15}
Z.~Wang, T.~Schaul, M.~Hessel, H.~Van~Hasselt, M.~Lanctot, and N.~De~Freitas,
  ``Dueling network architectures for deep reinforcement learning,'' in
  \emph{International Conference on Machine Learning {(ICML)}}, 2016, pp.
  1995--2003.

\bibitem{50}
A.~Tavakoli, F.~Pardo, and P.~Kormushev, ``Action branching architectures for
  deep reinforcement learning,'' \emph{arXiv preprint arXiv:1711.08946}, 2017.

\bibitem{51}
P.~A. {Lopez}, M.~{Behrisch}, L.~{Bieker-Walz}, J.~{Erdmann}, Y.~{Flötteröd},
  R.~{Hilbrich}, L.~{Lücken}, J.~{Rummel}, P.~{Wagner}, and E.~{Wiessner},
  ``Microscopic traffic simulation using sumo,'' in \emph{Proc. 21st Int. Conf.
  on Intel. Trans. Sys. (ITSC)}, 2018, pp. 2575--2582.





\end{thebibliography}
\end{document}